\documentclass[11pt,a4paper]{article}
\usepackage{jheppub}
\usepackage[T1]{fontenc}
\usepackage{lmodern}
\usepackage{booktabs,mathtools,microtype,mathrsfs,placeins}
\usepackage{flafter}
\newcommand{\caln}{\mathcal N}
\newcommand{\Ntwo}{\ensuremath{\mathcal N=2^*}}
\newcommand{\dd}{\mathrm d}
\newcommand{\cO}{\mathcal O}
\newcommand{\ent}{\mathcal S}
\newcommand{\press}{\mathcal P}
\newcommand{\energy}{\mathcal E}
\newcommand{\Puv}{P_{20}}
\newcommand{\ph}{p_h}
\newcommand{\PW}{\mathrm{PW}}
\title{PINNs for $\caln=2^*$ holography}
\author{Alex Buchel}
\affiliation{Department of Physics and Astronomy, University of Western Ontario, London, Ontario N6A 5B7, Canada}
\emailAdd{abuchel@uwo.ca}

\abstract{
Can a holographic theory be reconstructed from its equation of state?
We investigate this question using physics-informed neural networks in
the bosonic-mass sector of the Pilch--Warner model.  We first validate
the PINN approach against an analytic solution encoding the leading
high-temperature corrections to the conformal equation of state.  We
then reconstruct nonlinear black-brane geometries and the scalar
potential using thermodynamics, ultraviolet asymptotics, Ward
identities, and horizon regularity.  Independent solutions of the bulk
Einstein-scalar equations demonstrate that accurately reproducing
thermodynamics with a trained PINN does not guarantee an accurate
reconstruction of the scalar potential.  High-precision knowledge of
the high-temperature equation of state constrains nonlinear
interactions in the bulk scalar potential.  Fixing the cubic coupling
to its exact PW value, motivated by this high-temperature extraction,
reduces the maximum sampled error in the reconstructed scalar
self-interactions by a factor of $\sim 15$ compared with constraining
only the bulk scalar mass.  Adding knowledge of the bulk viscosity of
the gauge theory plasma further reduces this error by an additional
factor of $\sim 13$, with independent forward solutions reproducing
the bulk viscosity within $0.09\%$.
}

\hypersetup{pdftitle={PINNs for N=2* holography},pdfauthor={Alex Buchel}}
\begin{document}
\maketitle

\section{Introduction}
\label{sec:intro}
You wake up in the morning with the thought: how do I construct a
black brane in five dimensions with the equation of state and
hydrodynamics of water? The question is strange---water is
nonrelativistic, after all---but it touches on a serious question: how
robust is holography as a way of encoding strongly coupled matter?
There is a practical version of this question for the quark--gluon
plasma. We would really like to have a holographic dual that properly
accounts for its equation of state and hydrodynamics. Such a model
would open opportunities to address a vast array of
far-from-equilibrium questions for which comparably controlled tools
are presently scarce \cite{QGPreview,CheslerYaffe}. Matching
equilibrium data would not establish that all of those predictions are
correct; but it would give us a concrete model in which to ask the
questions.

This philosophy has a long history in bottom-up holography. Improved
holographic QCD and related Einstein--scalar constructions encode
running couplings, confinement and thermal physics in a suitable
gravitational action
\cite{IHQCD1,IHQCD2,IHQCDthermo,GubserNellore}. The difficult step is
to turn this philosophy into an inverse calculation: given an equation
of state, determine the bulk theory that produces it. The pioneering
work of Bea et al.\ \cite{Bea2024} made this step practical with
physics-informed neural networks (PINNs).
Its crucial contribution is to reconstruct a common bulk scalar
potential from a family of thermal states, rather than reconstructing
an isolated geometry. We build
directly on that idea.

Our purpose is a sequence of controlled experiments. We successively
provide more holographic information to the training, and compare the
result with a model whose potential is known independently. The
example is the Pilch--Warner (PW) construction \cite{PilchWarner}, one
of the best studied nonconformal examples of gauge/gravity
duality. The underlying scalar potential is known analytically,
while the finite-temperature backgrounds used to generate
the training data are high-precision numerical solutions.
 This distinction matters. We are not asking the network to reproduce a
closed-form nonlinear black brane. The dimension-two deformation, its
logarithmic boundary behaviour, and the relation between ultraviolet
and horizon data also introduce technical issues absent from the
dimension-three benchmark of Bea et al. Resolving these issues in a
controlled model should be useful well beyond this example.

We begin with a restricted class of inverse problems.
We consider homogeneous and isotropic states with no conserved
charge density, whose nonconformal equation of state has a single
intrinsic scale $m$ besides temperature.
We assume that the ultraviolet scaling dimension $\Delta$ of the
operator responsible for breaking conformal invariance is known,
and that its source $J$ is fixed in terms of the deformation scale,
$J\propto m^{4-\Delta}$.
The dimensionless deformation parameter is therefore
$J/T^{4-\Delta}\propto(m/T)^{4-\Delta}$. A description by one active bulk scalar is an
additional assumption, not a consequence of the number of scales. The
thermodynamic input is sufficiently precise that, at fixed source,
\begin{equation}
 \mathcal F=\energy-T\ent=-\press,\qquad
 \energy+\press=T\ent,\qquad \dd\energy=T\dd\ent
 \label{eq:thermo-identities}
\end{equation}
can be enforced consistently. Extensions to uncertain data, additional
scales and several operators are discussed at the end.
In the final experiment, we also assume knowledge of the
bulk viscosity to entropy density ratio, $\zeta/\ent$.

The experiments follow a simple progression. We first test the
geometry solver with a fixed potential, beginning with an exact
linearized benchmark.
We then reconstruct the potential from thermodynamics,
enforcing the holographic Ward identity, the first law,
and horizon regularity. Next we determine which nonlinear
ultraviolet interactions can actually be inferred from the
high-temperature data, and test the effect of fixing the cubic
coupling in the bulk scalar potential.
Finally, we incorporate bulk viscosity, providing an independent
transport constraint on the reconstructed scalar dynamics.

We focus on bulk viscosity because shear viscosity supplies no
analogous discrimination within the isotropic, two-derivative Einstein
class of the effective actions: $\eta/\ent=1/(4\pi)$ is temperature independent
\cite{BuchelLiuUniversal,BuchelUniversal}. Higher-derivative
extensions change this situation. The transport relations developed in
\cite{BuchelCremoniniEarly,BuchelGB,BuchelCharged} offer a route for
incorporating shear and other transport data in the same spirit.

This is the first paper in that research programme.
Our objective is not to compete with conventional numerical
integration when the bulk potential is known, but to determine
which physical inputs distinguish an accurate reconstruction
of thermodynamics from an accurate reconstruction of the
underlying theory. This distinction becomes essential when
working with phenomenological data of finite precision
and limited temperature range.

Section~\ref{sec:pw} reviews the PW model and establishes our
holographic dictionary.
Section~\ref{sec:highT} derives constraints on the bulk potential
from high-temperature thermodynamics.
Sections~\ref{sec:linear} and \ref{sec:direct} develop the
linearized and nonlinear direct PINNs.
Section~\ref{sec:inverse} presents the inverse reconstruction
and independent forward validation.
Section~\ref{sec:transport} incorporates bulk viscosity,
and section~\ref{sec:discussion} discusses limitations
and extensions.

\section{The PW laboratory and the thermodynamic dictionary}
\label{sec:pw}
\subsection{The field theory and its thermodynamics}
\label{sec:pw-physics}
Let us recall the field-theoretic description of the mass
deformation of $\mathcal N=4$ supersymmetric Yang--Mills theory
\cite{HoyosPaikYaffe}.
The $SU(N)$ theory contains a gauge field $A_\mu$, four adjoint
fermions $\psi_a$ and three adjoint complex scalars $\phi_i$.
Following the conventions reviewed in \cite{BuchelQuench}, the two
mass sources enter the Lagrangian density as
\begin{equation}
 \delta\mathcal L=-2\bigl(m_b^2\cO_2+m_f\cO_3\bigr),
 \label{eq:qft-deformation}
\end{equation}
where the bosonic and fermionic mass operators are
\begin{align}
 \cO_2&=\frac13\operatorname{Tr}
 \bigl(|\phi_1|^2+|\phi_2|^2-2|\phi_3|^2\bigr),
 \label{eq:qft-O2}\\
 \cO_3&=-\operatorname{Tr}\bigl(
 i\psi_1\psi_2-\sqrt2g_{\rm YM}\phi_3[\phi_1,\phi_1^\dagger]
 +\sqrt2g_{\rm YM}\phi_3[\phi_2^\dagger,\phi_2]
 +\mathrm{h.c.}\bigr)\nonumber\\
 &\quad+\frac23m_f\operatorname{Tr}
 \bigl(|\phi_1|^2+|\phi_2|^2+|\phi_3|^2\bigr).
 \label{eq:qft-O3}
\end{align}
Their ultraviolet dimensions are two and three, respectively.
The last term in $\cO_3$ is a source-dependent completion, so that
$m_f\cO_3$ includes a scalar mass contribution of order $m_f^2$.
For $m_b=m_f$, these terms give a common mass to the
$\caln=2$ hypermultiplet containing $\phi_{1,2}$ and $\psi_{1,2}$,
while the two mass contributions for $\phi_3$ cancel.
This is the supersymmetric $\Ntwo$ deformation of the action;
a thermal state does not preserve supersymmetry.
Here we instead set $m_f=0$ and $m_b\ne0$.
The resulting purely bosonic deformation is not supersymmetric,
and the negative $\phi_3$ mass term in \eqref{eq:qft-O2} remains.
There is no known homogeneous zero-temperature vacuum of the purely
bosonic deformation.

The PW solution provides the corresponding holographic dual
\cite{PilchWarner}. Its supersymmetric zero-temperature interpretation
in terms of a distribution of Coulomb-branch eigenvalues was developed
in \cite{BuchelPeetPolchinski}.\footnote{See also \cite{BuchelRussoZarembo}.} The five-dimensional truncation
contains scalars dual to both mass operators; setting the fermionic
scalar to zero consistently retains the dimension-two bosonic sector.
We use the established name PW for this known gravitational potential,
with the field-theory qualification just described.

Finite-temperature PW backgrounds and their thermodynamics have been
extensively studied \cite{BuchelLiuThermo,BuchelThermo}, as have their
hydrodynamic responses
\cite{BuchelHydro,BenincasaBuchelStarinets,BuchelPagnuttiBulk}.
Mass quenches and equilibration provide further applications of this
laboratory \cite{BuchelQuench,BuchelHellerMyers}.
The bosonic thermal branch exhibits a second-order transition at low
temperature. The specific heat diverges while the bulk viscosity
remains finite; the effective relaxation time entering the sound-channel
hydrodynamic expansion diverges \cite{BuchelRelaxation,BuchelCritical}.
The current numerical equation of state locates the critical point at
\begin{equation}
 \frac{T_c}{m_b}\simeq0.4303.
 \label{eq:PW-critical-temperature}
\end{equation}

\begin{figure}[htbp]
 \centering
 \includegraphics[width=\textwidth]{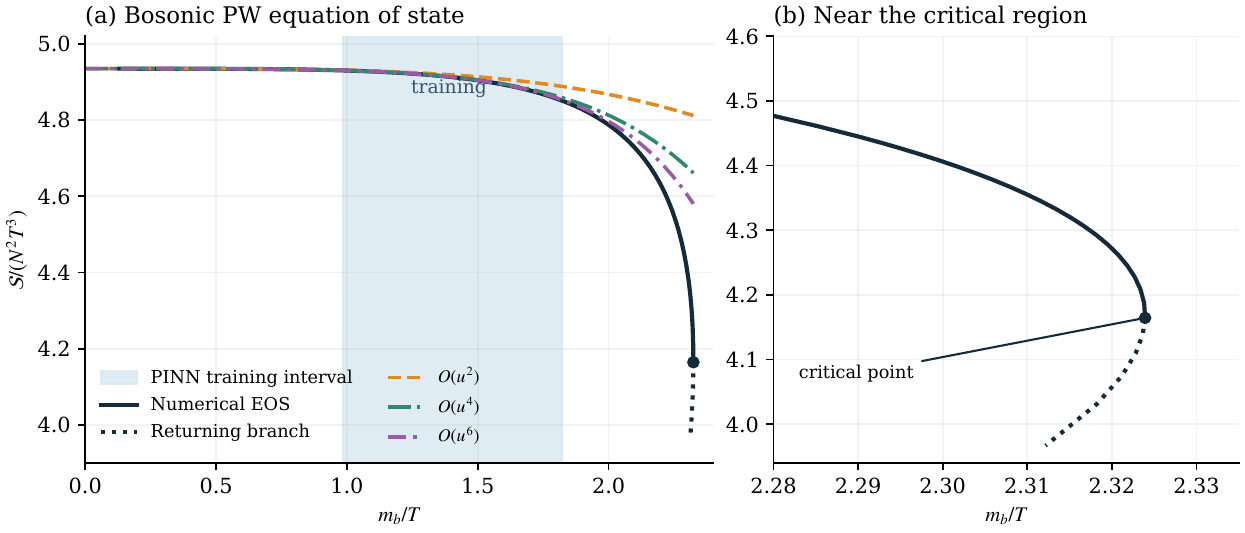}
 \caption{The numerical bosonic PW equation of state and its
 high-temperature approximations through $u^2$, $u^4$ and $u^6$.
 Panel (a) shows the full temperature range; shading marks the interval
 used for PINN training in section~\ref{sec:inverse}.
 Panel (b) resolves the critical region. The dot marks the critical
 point, and the dotted curve is the returning numerical branch.}
 \label{fig:PWentropyHT}
\end{figure}

Figure~\ref{fig:PWentropyHT} compares the numerical equation of state
of the bosonic PW plasma with successive high-temperature approximations.
The critical point and the temperature interval used for PINN training
in section~\ref{sec:inverse} are indicated. We denote the entropy
density by $\ent$ and plot $\ent/(N^2T^3)$,
which approaches $\pi^2/2$ in the conformal limit, and use
\begin{equation}
 u=\frac{1}{24\pi}\left(\frac{m_b}{T}\right)^2
 \label{eq:u-critical}
\end{equation}
as the high-temperature expansion parameter.
Remarkably, the high-temperature expansion retains information about
the critical point visible in figure~\ref{fig:PWentropyHT}a.
As we show in appendix~\ref{app:critical-extrapolation}, the known
square-root critical behaviour allows us to estimate its location
from the first few expansion coefficients.

Our inverse reconstruction focuses on a moderate-temperature
interval along the stable thermal branch
(the shaded region in figure~\ref{fig:PWentropyHT}a),
\begin{equation}
 1\lesssim\frac{m_b}{T}\lesssim2.
 \label{eq:training-mass-range}
\end{equation}
This interval is particularly useful for testing the
reconstruction methodology. The mass deformation is not
perturbatively small compared with temperature, so the
plasma exhibits appreciable departures from conformality.
At the same time, the interval remains away from the
critical region, where thermodynamic response functions
become singular.
We therefore test the reconstruction in a genuinely
nonconformal regime without introducing the additional
complications associated with critical phenomena.
Likewise, we avoid training in the high-temperature regime $T\gg m_b$,
where perturbative thermodynamics can be used directly to determine
the polynomial coefficients of the bulk scalar potential, as discussed
in appendix~\ref{app:highT}.

The two temperature regimes also suggest different choices of scale.
Setting $T=1$ is convenient for the high-temperature analysis: the
mass deformation becomes a small parameter, and successive terms in
the equation of state determine the polynomial coefficients of the
bulk scalar potential. This extraction requires no neural-network
training. As we explain in appendix~\ref{app:highT}, it allows us to
recover both the cubic and quartic interactions from the PW
thermodynamics.

For the appreciably nonconformal regime considered here, we instead
fix the scalar-source scale, $m=1$, and vary the temperature,
following Bea et al.~\cite{Bea2024}.
Here $m$ denotes the mass scale set by the scalar source; in our
normalization $m_b^2=\sqrt6\,m^2$.
This choice proved substantially more effective in our PINN
experiments.\footnote{Our earlier implementation with $T=1$ did not
achieve satisfactory reconstruction over this regime.} The two choices
describe the same dimensionless physics, but organize the numerical
problem differently. We therefore use temperature units for the
perturbative extraction of the potential coefficients and fixed-mass
units for training away from conformality.

\subsection{Holographic conventions and thermodynamic dictionary}
\label{sec:pw-dictionary}
Our action and field conventions are
\begin{equation}
 I=\frac{2}{\kappa_5^2}\int\dd^5x\sqrt{-g}
 \left[\frac14 R-\frac12(\partial p)^2-V(p)\right],
 \qquad \kappa_5^2=2,\qquad L_{\rm AdS}=1,
 \label{eq:action}
\end{equation}
with $p=\sqrt6\alpha$ in terms of the usual PW scalar. The reference potential is
\begin{align}
 V_\PW(p)&=-e^{-2cp}-2e^{cp},\qquad c=\frac{\sqrt6}{3},\label{eq:PWpot}\\
 V(p)&=-3-2p^2+v_3p^3+v_4p^4+\cO(p^5),\qquad
 v_3^\PW=\frac4{3\sqrt6},\quad v_4^\PW=-\frac13.
 \label{eq:Vuv}
\end{align}
The scalar mass squared $V''(0)=-4$ saturates the AdS$_5$ Breitenlohner--Freedman bound. We distinguish two computational problems:
\begin{itemize}
\item \textbf{Direct problem.} The scalar potential is specified. We
solve for the black-brane geometry and compute its thermodynamics.
In our benchmark, we supply the full PW potential and compare the
PINN solution with conventional numerical integration.
\item \textbf{Inverse problem.} The thermodynamic equation of state
is supplied. We reconstruct a scalar potential common to the thermal
family, together with the corresponding black-brane geometries.
Initially, only the constant, linear and quadratic terms in
\eqref{eq:Vuv} are fixed, encoding the AdS vacuum and the dimension
of the deforming operator. The nonlinear interactions are learned;
the known PW potential is used to assess the reconstruction.
Subsequent experiments supply additional information from
high-temperature thermodynamics and bulk viscosity.
\end{itemize}

We use ingoing Eddington--Finkelstein coordinates,
\begin{equation}
 \dd s^2=-\frac{a(r)}{r^2}\dd t_{\rm b}^2
 +\frac{s(r)^2}{r^2}\dd\boldsymbol{x}^{\,2}
 -\frac2{r^2}\dd t_{\rm b}\dd r,
 \label{eq:metric}
\end{equation}
with boundary $r=0$ and horizon $r=1$. The time coordinate $t_{\rm b}$ is distinct from a transformed radial coordinate $t$ used below. At unit source,
\begin{align}
 p&=r^2(\log r+\Puv)+r^3[-2k\log r-k(2\Puv+1)]+\cdots,\label{eq:UVp}\\
 s&=1+kr+\cdots,\nonumber\\
 a&=1+2kr+k^2r^2+r^4\!\left[A_{40}
 +\left(\frac2{27}-\frac89\Puv\right)\log r
 -\frac49\log^2r\right]+\cdots .
 \label{eq:UVmetric}
\end{align}
The coefficient $k=s'(0)$ retains the residual radial-coordinate
freedom needed to impose both our normalizations: unit scalar source
and a horizon at $r=1$. A scale transformation fixes the source,
while a shift of the inverse radial coordinate places the horizon at
$r=1$. This shift produces the linear terms proportional to $k$ in
the boundary expansions. Thus $k$ generally varies along the thermal
family even though the source and horizon position remain fixed.
Once these conventions are imposed, $k$ labels the direct solutions.
In the inverse problem, it is determined together with the geometry
rather than supplied as an additional thermodynamic observable.

Write $y=1-r$, $s=h+jy+\cdots$ and $p=\ph+\cdots$,
where $h=s(1)$, $j=-s'(1)$ and $\ph=p(1)$ are the horizon
coefficients. The source in \eqref{eq:UVp} has dimension two and
equals $m^2$ before setting $m=1$. Thermodynamic quantities below
are expressed in these source units unless their units are displayed
explicitly. The Hawking temperature and Bekenstein--Hawking entropy
density are
\begin{equation}
 T=-\frac{a'(1)}{4\pi},\qquad \ent=\pi h^3.
 \label{eq:TS}
\end{equation}
Holographic renormalization determines the boundary stress tensor
from the normalizable coefficients of the metric and scalar
\cite{BuchelQuench,BuchelKarapetyan}. In the convention used for
our thermodynamic data, the enthalpy density is
\begin{equation}
 \energy+\press=-A_{40}-\frac49\Puv^2
                       +\frac{2}{27}\Puv-\frac{13}{162}.
 \label{eq:renormalized-enthalpy}
\end{equation}
The thermodynamic relations $\mathcal F=\energy-T\ent$ and
$\mathcal F=-\press$, where $\mathcal F$ is the free-energy density,
give $\energy+\press=T\ent$. Consequently,
\begin{equation}
 A_{40}=-\frac49\Puv^2+\frac{2}{27}\Puv
          -\frac{13}{162}-T\ent.
 \label{eq:A40}
\end{equation}
This identity is built into the holographic dual: the bulk equations
equate the boundary enthalpy to the horizon quantity $T\ent$.

This is our \textbf{first} additional physics input beyond the implementation
of Bea et al.~\cite{Bea2024}. There, temperature and entropy are
imposed at the horizon, while the corresponding ultraviolet
coefficients emerge from training the geometry. We enforce
$\energy+\press=T\ent$ directly in the network parametrization.
The identity therefore holds exactly throughout training,
independently of the current bulk-equation residuals, and $A_{40}$
is eliminated as an independent trainable coefficient.

The trace Ward identity supplies a second algebraic constraint.
For a deformation by an operator $\mathcal O_\Delta$ of scaling
dimension $\Delta$, with source $\lambda_\Delta$ of mass dimension
$4-\Delta$, it takes the form \cite{BuchelQuench}
\begin{equation}
 \langle T^\mu{}_{\mu}\rangle
 = (\Delta-4)\lambda_\Delta\langle\mathcal O_\Delta\rangle+\mathcal A,
 \label{eq:generic-ward}
\end{equation}
where $\mathcal A$ is the conformal anomaly. In thermal equilibrium,
$\langle T^\mu{}_{\mu}\rangle=-\energy+3\press$.
For the bosonic deformation, $\Delta=2$, and holographic
renormalization expresses the scalar expectation value through $\Puv$.
The resulting interaction measure is
\begin{equation}
 \Delta_{\rm tr}\equiv\energy-3\press
                 =2\Puv+\frac12\log96.
 \label{eq:ward}
\end{equation}
Thus the input energy and pressure fix the scalar's normalizable
coefficient,
\begin{equation}
 \Puv=\frac12(\energy-3\press)-\frac14\log96,
 \label{eq:ward-P20}
\end{equation}
and \eqref{eq:A40} then fixes $A_{40}$. Imposing the trace Ward
identity exactly is our \textbf{second} additional physics input
relative to Bea et al.~\cite{Bea2024}. Together with the thermodynamic
identity above, it fixes both $\Puv$ and $A_{40}$ from the supplied
equation of state. Neither coefficient is therefore an independent
state parameter in the inverse reconstruction.
Equivalently, when solving the direct
problem, the thermodynamic quantities follow as
\begin{equation}
 \press=\frac14\left(T\ent-2\Puv-\frac12\log96\right),\qquad
 \energy=\frac14\left(3T\ent+2\Puv+\frac12\log96\right).
 \label{eq:energy-pressure}
\end{equation}

The constant $\tfrac12\log96$ in the interaction measure reflects
our scalar-source normalization and the renormalization convention
used for the thermodynamic data. Holographic renormalization permits
a finite local counterterm proportional to $m_b^4$
\cite{BuchelQuench,BuchelKarapetyan}. At fixed mass, this shifts the
energy density by a temperature-independent constant and the pressure
by its negative. Consequently, the interaction measure shifts, while
the enthalpy $\energy+\press=T\ent$ remains unchanged. The scalar
expectation value shifts consistently, preserving the trace Ward
identity.\footnote{The renormalization-scheme independence of the
trace Ward identity, with the stress tensor and scalar expectation
value renormalized consistently, was emphasized in
\cite{BuchelKarapetyan}, see equation (3.12).} We use the convention in which the additional
temperature-independent integration constant in the free energy is
zero, and retain this same convention throughout the reconstruction.

Regularity at a nonextremal horizon provides further local relations
between the geometry and the potential. The connection between the
horizon scalar derivatives and the derivatives of the potential was
emphasized in \cite{BuchelCremoniniEarly}, see in particular its
equation (A.26). Here it follows directly by inserting the regular
Taylor expansions into the bulk equations. Since $a(1)=0$ is a
simple zero, the scalar equation contains an apparent pole
proportional to $1/a$. Multiplying that equation by $r^2a$ and
taking $r\to1$ leaves
$a'(1)p'(1)-V'(\ph)=0$; regularity requires this coefficient to
vanish. The leading Einstein constraint similarly gives
$a'(1)(h+j)=\tfrac43hV(\ph)$. Using \eqref{eq:TS}, we obtain
\begin{equation}
 h+j=-\frac{hV(\ph)}{3\pi T},\qquad
 p'(1)=-\frac{V'(\ph)}{4\pi T}.
 \label{eq:horizon}
\end{equation}
Once the horizon scalar value and the potential are specified, these
slopes are fixed. In the inverse PINN we impose these relations
exactly for the potential at each stage of training. This is our
\textbf{third} additional physics input relative to
Bea et al.~\cite{Bea2024}.
For the PW potential \eqref{eq:PWpot}, the temperature in source
units is explicitly
\begin{equation}
 T=\frac{h}{3\pi(h+j)}
 \left[e^{-4\ph/\sqrt6}+2e^{2\ph/\sqrt6}\right].
 \label{eq:PW-horizon-temperature}
\end{equation}
The coefficients $h,j,\ph$ here belong to the unit-source geometry.
Thermodynamics therefore fixes $h$, $a'(1)$, $\Puv$ and $A_{40}$;
the horizon scalar value remains to be determined by the bulk
solution. The conventional PW mass satisfies $m_b^2=\sqrt6\,m^2$,
so $m=1$ is not the convention $m_b=1$.

\section{High-temperature constraints on the scalar potential}
\label{sec:highT}
The high-temperature region is perturbative and provides information
about the local potential before any inverse optimization. Consider
a generic scalar potential with the same ultraviolet mass,
\[
 V(p)=-3-2p^2+v_3p^3+v_4p^4+\cdots.
\]
In the unit-source convention, set
\begin{equation}
 z=\frac1{\pi^2T^2}=\frac{(m_b/T)^2}{\sqrt6\pi^2},\qquad
 R(z)=\frac{\ent}{\pi^4T^3}.
 \label{eq:zR}
\end{equation}
The perturbative calculation in appendix~\ref{app:highT} gives
\begin{equation}
 R=1-\frac12z^2+\frac23 c_{2}v_3z^3
 +(K_{40}+c_{3a}v_3^2+c_{3b}v_4)z^4+\cO(z^5),
 \label{eq:Reos}
\end{equation}
where $c_2,c_{3a},c_{3b}$ and $K_{40}$ are numerical coefficients
whose values are tabulated in appendix~\ref{app:highT}.
Fitting the cubic term in the high-temperature entropy expansion
thus determines $v_3$; the quartic term then determines $v_4$.
Fitting the numerical equation of state in the $T=1$ parametrization
recovers $v_3$ and $v_4$ with relative errors of $0.00091\%$ and
$0.37\%$, respectively; appendix~\ref{app:highT} details the extraction.

High-temperature thermodynamics thus provides additional input for
the inverse reconstruction: coefficients of the scalar potential
can be determined before training the networks. In the experiments
below, we fix the cubic coefficient to its exact PW value and leave
the higher interactions to be learned. This allows us to isolate
how knowledge of the cubic interaction improves the reconstruction.

\section{The linearized experiment}
\label{sec:linear}
\subsection{Network architecture and training}
A physics-informed neural network (PINN) represents an unknown
function by a neural network and determines its parameters by
minimizing the residuals of the differential equations
\cite{RaissiPINN}. We first explain this construction for a probe
scalar, before introducing the coupled nonlinear geometry.
Following section~\ref{sec:pw-physics}, we fix the source scale to
$m=1$ and vary the temperature.

At the linearized level, the geometry is the AdS$_5$--Schwarzschild
black brane, written in our radial gauge:
\begin{equation}
 s(r)=1+kr,\qquad
 a(r)=(1+kr)^2-\frac{(1+k)^4r^4}{(1+kr)^2}.
 \label{eq:linear-background}
\end{equation}
The bulk scalar propagates as a probe: we retain its linear equation
of motion and neglect its backreaction on the geometry.
The temperature is
\begin{equation}
 T=-\frac{a'(1)}{4\pi}=\frac{1+k}{\pi}.
 \label{eq:linear-temperature}
\end{equation}
Thus, at fixed source scale, $k\to\infty$ is precisely the
high-temperature limit $T\to\infty$. The scalar equation is
\begin{equation}
 p''+\left(\frac{a'}a+\frac{3k}{1+kr}-\frac3r\right)p'
       +\frac{4p}{r^2a}=0,
 \label{eq:linearODE}
\end{equation}
where $a(r)$ is given in \eqref{eq:linear-background}.
As $k$ increases, the temperature grows while the source scale
remains fixed. The scalar amplitude therefore decreases, making
its backreaction perturbatively small. Indeed, as we demonstrate
in the next subsection, the regular linearized solution gives
\begin{equation}
 \ph=-\frac{\pi}{4(1+k)^2}=-\frac{1}{4\pi T^2},
 \label{eq:linear-probe-amplitude}
\end{equation}
where $\ph\equiv p(1)$ is the scalar's horizon value.
The linearized problem thus describes the leading high-temperature
scalar profile of the nonlinear black-brane family.

We introduce $\nu_p(r,k)\equiv\partial_r p(r,k)$. The scalar field
$p$ and its derivative $\nu_p$ are represented by two separate
neural networks, which we call the \emph{scalar network} and the
\emph{derivative network}, respectively. The relation between the scalar and its
independently represented derivative is enforced through the loss
function. The two residuals are
\begin{align}
 C_p={}&p'-\nu_p,\nonumber\\
 E_p={}&r^2(1+kr)a\nu_p'
 +\big[-3ra(1+kr)+r^2(1+kr)a'+3r^2ak\big]\nu_p
 +4(1+kr)p.
 \label{eq:linear-residual}
\end{align}
Their simultaneous vanishing is equivalent to
\eqref{eq:linearODE}. Training adjusts the network parameters to
minimize these consistency and scalar-equation residuals.

Before introducing the networks, we separate the known logarithmic
boundary behaviour. The importance of explicitly separating the
boundary logarithms from the numerically determined fields was
emphasized in \cite{BuchelQuench}; see also \cite{CheslerYaffe}.
We subtract
\begin{equation}
 \frac{r^2\log r}{(1+kr)^2}
 \label{eq:b0}
\end{equation}
from $p$, and its radial derivative from $\nu_p$.
The denominator resums the powers of $kr$ multiplying the leading
boundary logarithm. The networks describe the remaining functions,
which have milder boundary behaviour; their first unsubtracted
logarithmic terms in the scalar field start at $r^6\log r$.

The scalar network and the derivative network both
receive the radial position $r$ and the state label $k$.
The networks are evaluated at a discrete set of radial positions,
called \emph{collocation points}. These are commonly chosen as
uniformly spaced points, randomly sampled points, or Chebyshev nodes.
Here we use Chebyshev nodes, which place more points near the
boundary and horizon. The networks themselves encode continuous
functions of $r$ and $k$.

Each network is a parametrized function constructed by composing
simple operations. Its inputs pass through three successive
\emph{hidden layers}, each containing 32 computational units,
conventionally called neurons. Every neuron forms a weighted sum
of the preceding layer's outputs, adds an adjustable constant
called a bias, and applies the hyperbolic tangent. This nonlinear
operation is the \emph{activation function}. Each neuron has its
own weights and bias. The layers are called hidden because their
outputs are intermediate quantities, rather than the supplied
inputs or the final network output. These intermediate quantities
are not values at additional radial grid points.

\begin{figure}[tbp]
 \centering
 \includegraphics[width=.97\textwidth]{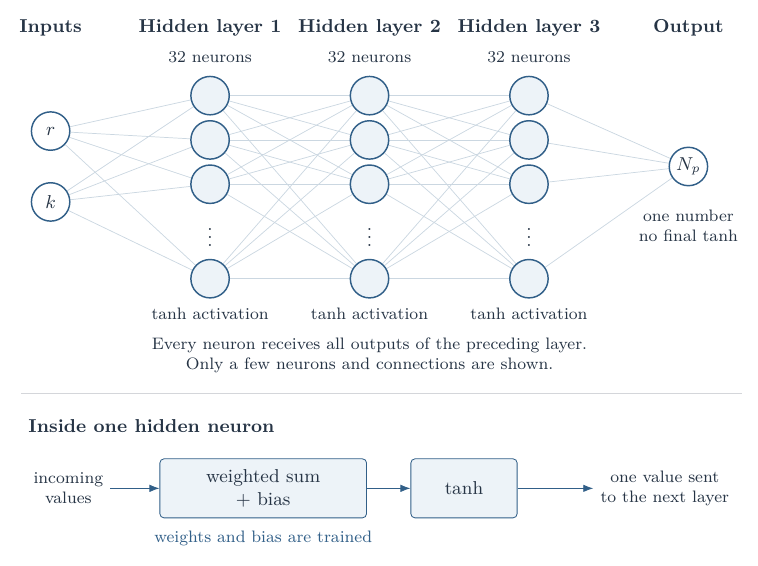}
 \caption{Structure of the scalar network. The two inputs pass
 through three hidden layers of 32 neurons and then a single output
 unit. Every neuron receives all outputs of the preceding layer;
 only a few neurons and connections are drawn. The inset shows the
 operation performed by each hidden neuron. The derivative network
 has the same structure with its own weights and biases.}
 \label{fig:linear-network-layers}
\end{figure}

After the third hidden layer, a weighted sum with an additional
bias produces one scalar output, without a final tanh. Training
adjusts the weights and biases. Thus, ``three layers of 32 tanh
neurons'' specifies the structure of the function being optimized.
We denote the scalar and derivative network outputs by $N_p(r,k)$
and $N_{\nu_p}(r,k)$, respectively.
Figure~\ref{fig:linear-network-layers} illustrates this construction;
\cite{PrinceDeepLearning} provides an elementary introduction.
The scalar and derivative networks have the same architecture but
separate parameters, with 4,482 trainable weights and biases in total.

We reconstruct the fields as
\begin{align}
 p(r,k)={}&\frac{r^2}{(1+kr)^2}
                  \big[\log r+N_p(r,k)\big],\nonumber\\
 \nu_p(r,k)={}&\frac{r\,[2\log r+1+kr]}{(1+kr)^3}
 +\frac{r}{(1+kr)^3}\Big[2N_p(0,k)\nonumber\\
 &\hspace{13mm}+r\big\{N_p(1,k)-(1+k)-2N_p(0,k)
                      +(1-r)N_{\nu_p}(r,k)\big\}\Big].
 \label{eq:linear-ansatz}
\end{align}
The first term in each expression supplies the explicit logarithmic
subtraction. The remaining factors impose the source normalization
and the regular horizon relation exactly, for any network weights.
In particular, the unknown endpoint coefficients are
\begin{equation}
 \Puv(k)=N_p(0,k),\qquad
 \ph(k)=\frac{N_p(1,k)}{(1+k)^2}.
 \label{eq:linear-network-endpoints}
\end{equation}
Training determines the network functions $N_p(r,k)$ and
$N_{\nu_p}(r,k)$. Their weights and biases encode the entire family
of solutions. In particular, $\Puv(k)$ and $\ph(k)$ follow from
the endpoint values in \eqref{eq:linear-network-endpoints}; they
are not independent fit parameters for each temperature, or
equivalently, each value of $k$. The appearance of the scalar
network's endpoint values in $\nu_p$ ensures compatible boundary
conditions; the derivative network determines its remaining radial
shape. The equality $p'=\nu_p$ must still be learned throughout the bulk.

\begin{figure}[tbp]
 \centering
 \includegraphics[width=.88\textwidth]{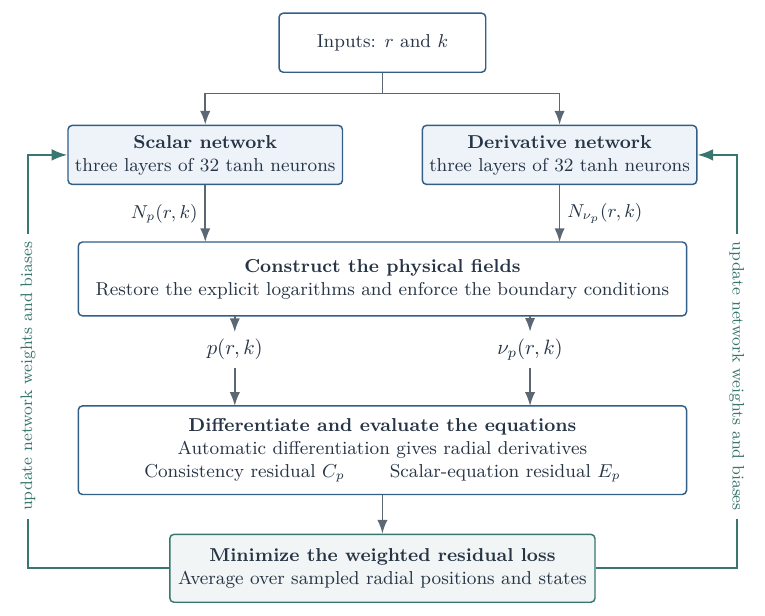}
 \caption{Training the linearized PINN. The scalar and derivative
 networks produce the two unknown functions used in
 \eqref{eq:linear-ansatz}. The explicit construction restores the
 logarithms and enforces the boundary conditions, using the scalar
 network's endpoint values for both fields. Radial derivatives
 enter the two equation residuals. Their weighted loss updates both
 networks; the exact bulk solution is used only for validation.}
 \label{fig:linear-pinn-training}
\end{figure}

Automatic differentiation supplies the radial derivatives of these
expressions. At the collocation points for each sampled thermal
state, we evaluate $C_p$ and $E_p$.
Training adjusts the weights of the scalar and derivative networks
jointly to minimize their residual loss, as illustrated in
figure~\ref{fig:linear-pinn-training}. Numerical scalar profiles
are not supplied as training targets.

The differential equations do not uniquely prescribe a loss function:
multiplying a residual by a nonzero function preserves its exact
solutions but changes how errors contribute during training.
Choosing this weighting is therefore part of constructing the PINN.
This issue is not specific to the PW model. In particular, a residual
can become small near a boundary simply because it contains a
vanishing radial prefactor, even when the unknown boundary
coefficients are incorrect. Such a region then contributes little
to the loss, and training has little incentive to improve it.

The appropriate weights depend on the equations and their boundary
asymptotics. Here, with the boundary behaviour already imposed,
$C_p=\cO(r^2)$ and $E_p=\cO(r^3)$. We therefore use
\begin{equation}
 \mathcal L_{\rm lin}=\frac12\left\langle
 \left(\frac{C_p}{r^2}\right)^2+
 \left(\frac{E_p}{r^3}\right)^2\right\rangle,
 \label{eq:linloss}
\end{equation}
where the brackets denote an average over the collocation points.
The quotients are evaluated after analytic cancellation of the
boundary powers, with finite limits at $r=0$; details are given in
appendix~\ref{app:linear}. This rescaling changes the relative
importance of errors during training while preserving the solutions
for which both residuals vanish.

We train at 17 logarithmically spaced values on $1\leq k\leq10$,
using 48 Chebyshev radial nodes and $1.5\times10^6$ Adam updates.
Adam is a gradient-based optimization algorithm which adjusts the
network weights using derivatives of the loss with respect to
those weights. The initial learning rate is $10^{-3}$, reduced by a
factor $0.985$ every 5,000 updates. Sixteen intermediate states
monitor interpolation. These and the subsequent nonlinear
calculations use float64 arithmetic.
The linearized training can be carried out on a single workstation
GPU: the $1.5\times10^6$ updates reported here took approximately
3 hours 41 minutes on an NVIDIA Quadro RTX 5000.

\subsection{Exact solution and validation}
\label{sec:linear-exact}
The regular solution with unit logarithmic source can be written
explicitly as
\begin{equation}
 p_{\rm ex}(r,k)=-\frac{\pi r^2}{4(1+kr)^2}
 {}_2F_1\!\left(\frac12,\frac12;1;
 1-\frac{(1+k)^4r^4}{(1+kr)^4}\right),
 \label{eq:exact}
\end{equation}
where ${}_2F_1$ is the Gauss hypergeometric function.
Its boundary and horizon limits give
\begin{equation}
 \Puv(k)=\log\frac{1+k}{2},\qquad
 \ph(k)=-\frac{\pi}{4(1+k)^2}.
 \label{eq:linear-exact-coeff}
\end{equation}
Appendix~\ref{app:linear} gives the derivation.
These formulas are reserved for validation: they supply neither
training targets nor criteria for selecting the trained network.
In particular, $\Puv(1)=0$ exactly, so we use an absolute error for
this coefficient near $k=1$.

The single-state $k=1$ experiment illustrates the importance of
residual weighting. The unweighted loss reached about
$6.82\times10^{-9}$, yet the network gave $\Puv=-0.114(8)$ and
a $3.55\%$ horizon-scalar error. With the rescaled loss
\eqref{eq:linloss}, the errors improved to $\Puv=-7.85\times10^{-5}$
and $0.0116\%$ at the horizon. A small numerical loss alone is
therefore insufficient to assess the accuracy of the solution.

For the continuous family, the maximum horizon error over 129
evaluation states is $0.1683\%$, and the largest absolute error
in $\Puv$ is $4.53\times10^{-4}$. Trained and intermediate states
have comparable errors. The network thus learns both the radial
profile and its dependence on temperature at sub-percent accuracy
for the horizon coefficient.

\subsection{Effect of higher-order logarithmic subtraction}
To test whether removing more of the boundary logarithms improves
training, we replace \eqref{eq:b0} by
\begin{equation}
 \frac{r^2\log r}{(1+kr)^2}
 \left[1+\frac{(1+k)^4r^4}{4(1+kr)^4}\right].
 \label{eq:b1}
\end{equation}
This also removes the logarithmic terms beginning at order
$r^6\log r$; the first remaining logarithm in the scalar field
appears at order $r^{10}\log r$. We adjust the factors enforcing
the horizon boundary condition consistently and keep the network
architecture and training procedure unchanged.
Table~\ref{tab:linear-sub} shows that the additional subtraction
does not improve the recovered horizon coefficient.
\begin{table}[tb]
\centering
\begin{tabular}{lrr}
\toprule
Sampled maximum error & Leading subtraction & Extended subtraction\\
\midrule
Trained $\Puv$, absolute & $4.533\times10^{-4}$ & $4.901\times10^{-4}$\\
Intermediate $\Puv$, absolute & $3.359\times10^{-4}$ & $3.728\times10^{-4}$\\
Trained $\ph$, percent & $0.1665$ & $0.9970$\\
Intermediate $\ph$, percent & $0.1623$ & $0.3465$\\
All evaluated $\ph$, percent & $0.1683$ & $0.9970$\\
$1\leq k\leq5$: $\ph$, percent & $0.08047$ & $0.142(6)$\\
\bottomrule
\end{tabular}
\caption{Comparison of the leading and extended logarithmic
subtractions. The last row restricts the evaluation to moderate
$k$. Each column reports one training run.}
\label{tab:linear-sub}
\end{table}

Automatic differentiation and an independent Chebyshev-based
calculation give consistent derivatives. Automatic differentiation
applies the chain rule to the represented functions; it does not
assume that the target solution is analytic. In these experiments,
the leading subtraction and residual rescaling are sufficient for
accurate reconstruction, and removing further logarithms offers
no improvement. We therefore retain the leading subtraction in
studying the nonlinear problem at moderate temperature.

\section{Learning the nonlinear geometry at fixed potential}
\label{sec:direct}
\subsection{Six networks and a resummed boundary expansion}
We now keep \eqref{eq:PWpot} fixed and learn $p,s,a$ together with three independently represented derivatives of the transformed fields introduced below. The derivative networks represent $v_p,v_s,v_a$; their physical radial counterparts, $\nu_p,\nu_s,\nu_a$, follow from \eqref{eq:slopes}. There are six separate $2$--$32$--$32$--$32$--$1$ tanh networks, with 13,446 trainable parameters. This retains the geometry capacity of the linear experiment for each unknown function. Boundary coefficients are endpoint evaluations of these networks, not additional per-state tables.

For each fixed $k$, define
\begin{equation}
 K=1+k,\quad w=1+kr,\quad t=\frac{Kr}{w},\quad
 S=\frac{s}{w},\quad A=\frac{a}{w^2},\quad \epsilon=K^{-2},\quad\mu=K^{-4}.
 \label{eq:normalized-fields}
\end{equation}
The rescaled metric function $S$ is distinct from the entropy density $\ent$. The three slope networks represent $v_p,v_s,v_a$, intended to be $p_t,S_t,A_t$. Their physical radial counterparts are
\begin{equation}
 \nu_p=\frac K{w^2}v_p,\qquad
 \nu_s=kS+\frac K w v_s,\qquad
 \nu_a=2kwA+Kv_a.
 \label{eq:slopes}
\end{equation}
Let $b=\Puv-\log K$ and $B=\log t+b$. The leading nonlinear bases are
\begin{align}
 p_U&=\epsilon t^2B+\beta\epsilon^2t^4\left(B^2-2B+\frac32\right),\qquad
 \beta=\frac34v_3,\nonumber\\
 S_U&=1+\mu t^4\left[-\frac29B^2+\frac{B}{27}-\frac{13}{324}\right],\nonumber\\
 A_U&=1+\mu t^4\left[-\frac49B^2+\frac{2B}{27}
                         -\frac{13}{162}-T\ent\right],\label{eq:bases}\\
 p&=p_U+\epsilon t^5H_p,\qquad
 S=S_U+\mu t^5H_s,\qquad
 A=A_U+\mu t^5H_a.\nonumber
\end{align}
For PW, $\beta=\frac{\sqrt6}{6}$. These expressions include smooth partners of the logarithmic subtraction and resum the powers of $kr$. They incorporate the known ultraviolet terms through order $r^4$. Higher logarithms remain; complete analyticity of the remainders is neither required (as explored in table~\ref{tab:linear-sub}) nor claimed. A useful explicit form of the logarithmic pieces is given in appendix~\ref{app:subtraction}.

Four state functions determine the free endpoint data:
\begin{align}
 b&=N_p(0,k),&J&=N_p(1,k),&\eta_h&=N_s(1,k),&\chi_h&=N_{v_s}(1,k),\nonumber\\
 \Puv&=\log K+b,&\ph&=\epsilon J,&h&=Ke^{\mu\eta_h},&h+j&=e^{\mu\chi_h}.
 \label{eq:direct-state}
\end{align}
Exponentials ensure positivity of $h$ and $h+j$. The metric horizon value, regularity and the thermodynamic dictionary determine the endpoint slopes and $T\ent$. The three consistency losses remain necessary: assigning a regular independent slope does not alone make the derivative of the field network regular.

\subsection{Equations and loss normalization}
With $D=S-tv_s$, the six residuals are
\begin{equation}
 C_p=p_t-v_p,\qquad C_s=S_t-v_s,\qquad C_a=A_t-v_a.
 \label{eq:consistency}
\end{equation}
\begin{align}
 E_p&=t^2SA(v_p)_t+(3t^2Av_s+t^2Sv_a-3tSA)v_p-SV'(p),\label{eq:Ep}\\
 E_s&=(v_s)_t+\frac23Sv_p^2,\label{eq:Es}\\
 E_a&=tSDv_a+\frac23t^2S^2Av_p^2-2AD(S+D)-\frac43S^2V(p).
 \label{eq:Ea}
\end{align}
The last equation is the first-order Einstein constraint. An unused second-order equation for $a$ supplies an independent numerical diagnostic of the approximate solution. These polynomial forms avoid divisions by learned metric fields.

The normalized residuals $\widehat R_\alpha$ divide \eqref{eq:consistency}--\eqref{eq:Ea}, in that order, by
\begin{equation}
 \epsilon t^4,\quad\mu t^4,\quad\mu t^4,\quad
 \epsilon t^5(1-t^4),\quad\mu t^3,\quad\mu t^5(1-t^4).
 \label{eq:divisors}
\end{equation}
As in the linear problem, we cancel known terms analytically before division. The ultraviolet powers remove suppression caused by the imposed expansion. The horizon factors remove zeros already enforced by regularity. Numerically dividing a tiny, cancellation-prone raw expression by a tiny divisor is not an equivalent implementation in finite precision.

The algebraic thermodynamic constraints of
section~\ref{sec:pw-dictionary} hold by construction.
The first law (FL) along the family supplies an independent differential
constraint. Its inclusion in the training loss is our \textbf{fourth}
additional physics input relative to the implementation of
Bea et al.~\cite{Bea2024}. For $\tau=\log K$,
\begin{align}
 F_{\rm FL}&=\partial_\tau\Puv+\frac32\partial_\tau(T\ent)
              -6T\ent\partial_\tau\log h\nonumber\\
 &=1+b_\tau+\frac32\partial_\tau(T\ent-K^4)-6(T\ent-K^4)
       -6\mu T\ent(\eta_{h,\tau}-4\eta_h).
 \label{eq:firstlaw}
\end{align}
The second line cancels large conformal contributions explicitly. Since
$\press_\tau-\ent T_\tau=-F_{\rm FL}/2$, its vanishing is the first law. The direct objective is the loss function
\begin{equation}
 \mathcal L_{\rm dir}=\frac17\left[
 \sum_{\alpha=1}^{6}\langle\widehat R_\alpha^2\rangle_{r,k}
 +\langle F_{\rm FL}^2\rangle_k\right].
 \label{eq:directloss}
\end{equation}
State and node weights are equal. In particular, there is no sliding state weight favouring one part of the $k$ interval. The brackets are arithmetic node means, not invariant radial integrals. Hard identities are not counted as independently learned successes.

\subsection{Direct results and the location of error}
The baseline uses 17 states on $1.69(9)\lesssim k\lesssim3.9982$ and 64 fixed open Chebyshev nodes. With the optimizer schedule used above, a twelve-hour run on a Quadro RTX 5000 completed 1,166,422 updates. The checkpoint was selected by training loss; benchmark labels were not used to select it. Table~\ref{tab:direct} reports its errors against conventional numerical solutions.
\begin{table}[tb]
\centering
\begin{tabular}{lrr}
\toprule
Maximum error & 17 trained states & 167 untrained reference states\\
\midrule
$\Puv$, absolute & $2.56(9)\times10^{-4}$ & $9.11(4)\times10^{-5}$\\
$\ph$, percent & $0.02154$ & $0.007981$\\
$T$, percent & $0.0001102$ & $0.00002268$\\
$\ent$, percent & $0.0006672$ & $0.0002403$\\
$\press$, percent & $0.0002605$ & $0.00009080$\\
$\energy$, percent & $0.0008705$ & $0.0003185$\\
$R-1$, percent & $0.06235$ & $0.02472$\\
\bottomrule
\end{tabular}
\caption{Nonlinear direct baseline. The untrained reference subset covers only $2.20(3)\leq k\leq3.9881$, so its smaller errors do not establish cold-end interpolation accuracy. $R-1$ isolates the nonconformal entropy signal.}
\label{tab:direct}
\end{table}

The dominant errors occur near the lower end of the state range. The thermodynamic accuracy is not simply dilution by a large conformal contribution: the nonconformal entropy correction is also recovered well below one percent. We use absolute errors for $\Puv$ near its zero.

Evaluating the residuals on a finer radial grid reveals errors missed
by the training points: the loss increases from approximately $10^{-7}$
to $3.2\times10^{-6}$, with similar values at trained and intermediate
temperatures. About $99.4\%$ of the radial residual loss comes from
$r<0.1$, predominantly from the metric equations. The main limitation
is therefore radial resolution near the boundary. This motivates
testing whether moving the collocation points during training improves
the solution, as we now explore in the next subsection.

\subsection{Moving the collocation points}
Three continuations start from the same final baseline checkpoint, with the same optimizer state and exactly 50,000 additional updates. The control leaves the nodes fixed. A jitter experiment uses one common phase per update,
\begin{equation}
 r_j=\frac{1-\cos[\pi(j+1/2+\delta)/64]}2,
 \qquad \delta\sim U[-1/4,1/4],\quad j=0,\ldots,63.
 \label{eq:jitter}
\end{equation}
A third experiment maps the usual open Chebyshev nodes $x_j$ to a fixed grid,
\begin{equation}
 r_j=x_j\left[1-\frac12(1-x_j)^2\right].
 \label{eq:uvgrid}
\end{equation}
This moves points toward the boundary while preserving leading horizon spacing. Neither experiment changes the state weights. Equal weights on the displaced nodes do change the effective radial sampling measure, so the experiment does not isolate aliasing from weighting.
\begin{table}[tb]
\centering
\begin{tabular}{lrrr}
\toprule
Common 127-node audit & Fixed & Phase jitter & UV-biased fixed\\
\midrule
Loss at trained states & $3.13(0)\times10^{-6}$ & $2.53(1)\times10^{-6}$ & $1.10(4)\times10^{-6}$\\
Loss at intermediate states & $3.05(3)\times10^{-6}$ & $2.46(0)\times10^{-6}$ & $1.07(7)\times10^{-6}$\\
UV $\widehat E_s$ RMS & $5.10(6)\times10^{-3}$ & $3.92(4)\times10^{-3}$ & $2.70(8)\times10^{-3}$\\
UV $\widehat E_a$ RMS & $8.75(9)\times10^{-3}$ & $7.46(6)\times10^{-3}$ & $5.04(4)\times10^{-3}$\\
\bottomrule
\end{tabular}
\caption{Final-step comparison after matched continuations. All values use the unchanged ordinary validation grid, not each experiment's different training measure. Residual RMS values are not percentage errors.}
\label{tab:collocation}
\end{table}

Jitter reduces dense loss by about $19\%$ relative to fixed continuation. The static UV-biased grid reduces it by about $65\%$, with a similar gain at intermediate states. Some smaller bulk and horizon components become worse, so this is not uniform improvement of every equation. Thermodynamic errors remain small: the best UV-biased checkpoint has maximum trained-state errors of $0.000109\%$ in $T$ and $0.000851\%$ in $\energy$. The evidence supports improving radial coverage where measured residuals concentrate. It does not justify enlarging the architecture or introducing state weights without further diagnosis. We retain the UV-biased grid for the inverse experiments. These are matched continuations of one model, not an ensemble study.

For context, Bea et al.\ report maximum squared field discrepancies $7\times10^{-6}$ and $2.5\times10^{-4}$ in two fixed-potential tests after five million updates \cite{Bea2024}. Their square roots are about $0.00265$ and $0.0158$. These are field-amplitude discrepancies, not our endpoint-coefficient percentages; the different models, state ranges and diagnostics preclude a direct performance ranking.

To test interpolation at the cold end, we compare with independent conventional solutions at ten untrained values $k=1.725,1.775,\ldots,2.175$. The UV-biased direct network gives maximum errors $3.65\times10^{-4}$ in $\Puv$ (absolute), $0.0291\%$ in $\ph$, $0.00118\%$ in $\energy$, and $0.0846\%$ in $R-1$. These comparisons use independent conventional solutions without retraining. They confirm sub-percent accuracy at these cold intermediate states, while retaining the distinction from the original reference subset in table~\ref{tab:direct}.

\section{Joint reconstruction of geometry and potential}
\label{sec:inverse}
\subsection{Coupling the geometry and potential networks}
The inverse problem contains two groups of unknown functions: the thermal family of geometries and one potential common to all states. We retain the six geometry networks and introduce
\begin{equation}
 V_\psi(p)=-3-2p^2+p^3\bigl[v_3+pQ_\psi(p)\bigr],
 \label{eq:Vnet}
\end{equation}
where $Q_\psi$ is a scalar-input MLP (a multilayer perceptron) with four hidden layers of width 16 and SiLU activation. In the baseline B0, $v_3$ is a trainable scalar initialized to zero; the final layer of $Q_\psi$ has small random, nonzero weights. The known exponential PW potential is absent from the training objective. The ultraviolet conditions are exact by construction, and $v_4=Q_\psi(0)$ remains free.

\begin{figure}[tbp]
 \centering
 \includegraphics[width=.98\textwidth]{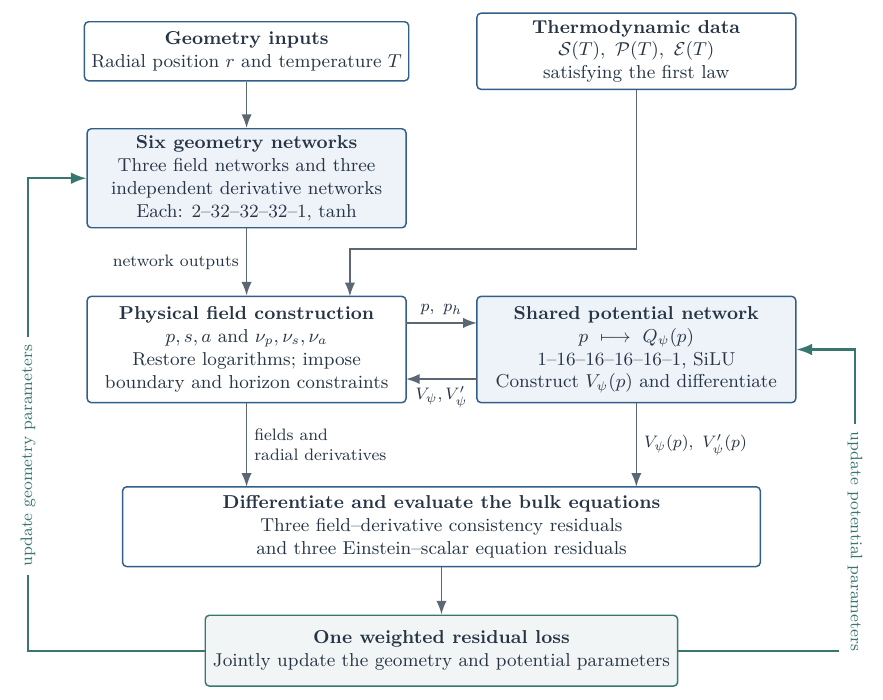}
 \caption{Joint training of the geometry and potential networks.
 The six geometry networks have three hidden layers of 32 tanh neurons
 each; the shared potential network has four hidden layers of 16 SiLU
 neurons. The six geometry networks contain 13,446 trainable weights
 and biases, and the potential MLP contains 865. Including the free
 cubic coefficient $v_3$, B0 has 14,312 trainable parameters; B1 and
 B2 have 14,311 each. The scalar field and its horizon value enter the potential
 network, whose potential and derivative also determine the horizon
 constraints. Automatic differentiation supplies the derivatives in
 the bulk equations. A single loss updates both groups of networks.
 The supplied thermodynamics satisfies the first law; the inverse
 loss has no additional first-law penalty.}
 \label{fig:inverse-network}
\end{figure}

Figure~\ref{fig:inverse-network} shows how the two groups of networks
interact. The six geometry networks receive the radial position $r$
and temperature label $T$. Together with the subtraction and boundary
conditions, their outputs encode $p,s,a$ and the three independently
represented radial derivatives. The potential network receives the
scalar value $p$, rather than $r$ or $T$, and returns $Q_\psi(p)$;
\eqref{eq:Vnet} then constructs the potential. The same potential
network is used at every temperature. Its layers perform the operations
illustrated in figure~\ref{fig:linear-network-layers}, with the
activation function replaced by $\operatorname{SiLU}(x)=x/(1+e^{-x})$.
The thermodynamic data fix boundary and horizon quantities as described
in the next subsection. The learned potential also enters the horizon
regularity relations and the boundary subtraction, so the two groups
are coupled through both the field construction and the equations.

At each collocation point, the geometry networks supply $p$, and the potential network evaluates $V_\psi(p)$ and $V_\psi'(p)$. Both groups of networks are optimized jointly through the Einstein--scalar residuals. Differentiation of the loss retains the full dependence on both groups, including the dependence through $V_\psi''$, horizon regularity and the ultraviolet subtraction.

The geometry and potential widths follow Bea et al., as does joint training \cite{Bea2024}. We omit their Gaussian localization layer, use one temperature coordinate on our chosen branch, and train six independent equations while auditing the redundant metric equation for $a''$. Our boundary subtraction and residual normalization address the dimension-two logarithms. The present narrow scalar interval supplies no evidence that a specialized localization layer is necessary; this remains testable if broader field coverage is required.

The subtraction must follow the potential being learned. In particular, $\beta=3v_3/4$ in \eqref{eq:bases} is variable in B0. Fixing this coefficient to its PW value while declaring $v_3$ unknown would leak the answer through the boundary expansion. Numerically stable evaluation also uses the explicit remainders
\begin{equation}
 \frac{V+3+2p^2}{p^3}=v_3+pQ_\psi(p),\qquad
 \frac{V'+4p-3v_3p^2}{p^3}=4Q_\psi(p)+pQ_\psi'(p),
 \label{eq:Vremainders}
\end{equation}
instead of subtracting nearly equal quantities near $p=0$.

\subsection{Thermodynamic boundary data and the remaining unknowns}
The input rows are $(T,\ent,\press,\energy)$ satisfying \eqref{eq:thermo-identities}. The network inputs are $r$ and $\log T$, with the temperature expressed in units of the source scale, $m=1$. We use a single-valued thermal branch; $T$ would not be a sufficient global label for a multivalued equation of state without branch information. The data determine
\begin{align}
 h&=(\ent/\pi)^{1/3},\qquad
 \Puv=\frac12(\energy-3\press)-\frac14\log96,\nonumber\\
 A_{40}&=-\frac49\Puv^2+\frac{2}{27}\Puv-\frac{13}{162}-T\ent.
 \label{eq:data-endpoints}
\end{align}
In the inverse problem, the temperature labels the thermal state,
while the gauge coefficient $k(T)$ is learned together with the geometry:
\begin{equation}
 k(T)=(h-1)\exp\left[\frac{N_a(0,T)}{h^4}\right],\quad
 K=1+k,\quad b=\Puv-\log K,\quad
 \ph=K^{-2}N_p(1,T).
 \label{eq:kappa}
\end{equation}
This parametrization is restricted to the moderate branch where $h>1$ and $k>0$. Its initialization near $h-1$ does not use the true $k$ values. No additional state network is introduced.

For given $\ph$ and the current potential, horizon regularity determines
\begin{equation}
 d_h=h+j=-\frac{hV_\psi(\ph)}{3\pi T},\qquad
 v_{a,h}=-\frac{4\pi T}{K},\qquad
 v_{p,h}=\frac{V_\psi'(\ph)}{v_{a,h}},\qquad
 v_{s,h}=\frac hK-d_h.
 \label{eq:inverse-horizon}
\end{equation}
Both network groups remain connected through these expressions.
After incorporating the four additional physics inputs described above,
the only remaining adjustable physical endpoint parameter at each
temperature is the horizon scalar value, $\ph=p(1)$. The gauge
coefficient $k(T)$ and the bulk profiles are learned together with it.
As we show in section~\ref{sec:transport}, incorporating bulk-viscosity
data, together with the high-temperature scalar asymptotics, determines
this remaining horizon parameter. This is our \textbf{fifth} and final
additional physics input relative to Bea et al.~\cite{Bea2024}.
The inverse objective is the loss function
\begin{equation}
 \mathcal L_{\rm inv}=\frac16\sum_{\alpha=1}^6
 \langle\widehat R_\alpha^2\rangle_{r,T}.
 \label{eq:inverseloss}
\end{equation}
It retains the normalized residuals, equal state weights and fixed UV-biased nodes. Thermodynamic consistency is checked in preparing the input. Since the thermodynamics is now supplied and imposed, an additional first-law loss would not constrain a new trainable physical quantity. This differs from the direct problem, where the thermodynamic trajectory itself is learned. Similarly, agreement with the supplied $\Puv$ and $A_{40}$ is not a prediction. Independent outputs include the potential, $k$, $\ph$, and the consistency of the inferred bulk solution.

\subsection{Thermodynamics alone: B0}
The baseline trains for 200,000 updates on 17 moderate-temperature states. Selection uses equation residuals on independent radial points at training states, not agreement with the PW answer. The resulting ultraviolet coefficients are
\begin{equation}
 v_3^{\rm B0}=0.512(9),\qquad v_4^{\rm B0}=-2.91(6).
 \label{eq:B0coeff}
\end{equation}
The cubic coefficient is low by $5.77\%$ and the quartic is far from $-1/3$. Yet the maximum absolute potential error on the sampled scalar interval, approximately $-0.143(7)\leq p\leq0$, is only $6.23\times10^{-6}$. On the common interval used in figure~\ref{fig:potential-reconstruction}(a), this maximum is $5.97\times10^{-6}$ (blue B0 curve). A plot of the full potential would largely display the supplied cosmological constant and mass. To expose the unknown interaction we therefore compare
\begin{equation}
 U(p)=\frac{V(p)+3+2p^2}{p^3},\qquad U(0)=v_3.
 \label{eq:U}
\end{equation}
The maximum sampled relative $U$ error is $5.76(8)\%$, shown by the blue B0 curve in figure~\ref{fig:potential-reconstruction}(b), or $3.85(2)\%$ if $|p|<0.02$ is excluded. The maximum training-state horizon-scalar error is $0.106(4)\%$.

One must still ask whether the learned potential actually produces the target thermodynamics. Evaluating the EoS-conditioned geometry network cannot answer that question, since its endpoint thermodynamics is an input. We freeze $V_\psi$, discard the geometry networks, and solve the conventional boundary-value problem with only temperature and the unit source specified. Entropy, pressure, energy, gauge coefficient and horizon scalar are then outputs. The audit uses the 17 training temperatures and ten additional cold-end temperatures corresponding to reference labels $k=1.725$--$2.175$. Repeated endpoint cutoffs and an exact-PW control resolve the numerical error of the forward calculation.

The forward B0 potential reproduces all three extensive thermodynamic densities at about $10^{-3}\%$ or better on these samples. Even the nonconformal entropy signal has maximum relative error only $0.254\%$, as shown by the blue B0 curve in figure~\ref{fig:eos-reconstruction}. Thus this is a successful thermodynamic reconstruction with imperfect interaction recovery. It demonstrates finite-precision ambiguity on the tested state and scalar intervals. It does not prove that the exact equation of state over every temperature admits inequivalent potentials with exactly identical observables.

\subsection{A known cubic interaction: B1}
The next experiment fixes $v_3=4/(3\sqrt6)$ in \eqref{eq:Vnet}, retaining the free MLP remainder, geometry architecture, moderate states, sampling and 200,000-update budget. This is the known-coefficient control motivated by section~\ref{sec:highT}; the exact coefficient is an additional input and is not counted as a recovered output. The selected checkpoint is the last update. Table~\ref{tab:inverse} compares the reconstructions.
\begin{table}[tb]
\centering
\begin{tabular}{lrr}
\toprule
Inverse diagnostic & B0: free $v_3$ & B1: fixed $v_3$\\
\midrule
Maximum relative $U$ error (\%) & 5.76(8) & 0.390(8)\\
Maximum relative $U$ error, $|p|\geq0.02$ (\%) & 3.85(2) & 0.353(5)\\
Maximum absolute $V'$ error & $5.53\times10^{-4}$ & $1.61\times10^{-4}$\\
Maximum training-state $\ph$ error (\%) & 0.106(4) & 0.01589\\
Free $v_4$ & $-2.91(6)$ & $-2.24(2)$\\
Common 193-node audit loss & $1.414\times10^{-6}$ & $1.744\times10^{-6}$\\
\bottomrule
\end{tabular}
\caption{Potential errors are evaluated on the common B0 scalar interval. The interaction improves despite a slightly larger aggregate equation loss. The quartic coefficient remains inaccurate.}
\label{tab:inverse}
\end{table}

We find that the improvement is physically useful: the maximum interaction error falls below $0.4\%$ (orange B1 curve in figure~\ref{fig:potential-reconstruction}(b)), and the horizon trajectory also improves. It is not accompanied by a uniformly smaller residual loss. About $86\%$ of the fixed-cubic dense loss remains in $r<0.1$. Ranking reconstructions solely by one scalar loss would obscure the result.

The independent forward comparison in table~\ref{tab:forward} uses precisely the same temperatures and numerical controls for B0 and B1. The largest error in the nonconformal entropy correction improves by a factor of about 11.4 at training temperatures and 4.2 at temperatures withheld from NN training; compare the orange B1 and blue B0 curves in figure~\ref{fig:eos-reconstruction}. The exact-PW control has maximum error $2.22\times10^{-5}\%$ in this correction, well below the reconstruction errors. A cold forward solution extends about $5.1\times10^{-6}$ beyond B1's sampled inverse scalar interval at the horizon.
\begin{table}[tb]
\centering
\begin{tabular}{lrrrr}
\toprule
&\multicolumn{2}{c}{Training temperatures}&\multicolumn{2}{c}{Additional temperatures}\\
Observable & B0 & B1 & B0 & B1\\
\midrule
$\ent$ & 0.000813(2) & 0.000384(9) & 0.000791(2) & 0.000221(7)\\
$\press$ & 0.000319(9) & 0.00006553 & 0.000253(1) & 0.00006611\\
$\energy$ & 0.00101(7) & 0.000527(0) & 0.000995(7) & 0.000307(5)\\
$\energy-3\press$ & 0.0203(8) & 0.0122(2) & 0.0185(9) & 0.007477\\
$R-1$ & 0.253(8) & 0.0222(6) & 0.0580(6) & 0.0137(5)\\
$(\energy-3\press)/(T\ent)$ & 0.0198(8) & 0.0118(4) & 0.0178(0) & 0.007256\\
\bottomrule
\end{tabular}
\caption{Maximum relative forward errors in percent. Only the frozen learned potential, temperature and source are supplied to the forward solver. The geometry networks and reference endpoint values are discarded.}
\label{tab:forward}
\end{table}

The free quartic coefficient $v_4$ remains far from its PW value in both experiments. Although the high-temperature analysis in appendix~\ref{app:highT-recovery} recovers this coefficient, we deliberately leave it unconstrained to test what bulk-viscosity information adds. The next experiment uses this additional observable to determine the remaining physical endpoint parameter, $p_h=p(1)$.

\section{Bulk viscosity as an input for the horizon scalar}
\label{sec:transport}
\subsection{The scalar normalization and the missing integration constant}
The Eling--Oz relation \cite{ElingOz,BuchelEO} connects bulk viscosity to the change of horizon scalars along the equilibrium family. Its normalization must be handled carefully. Rewriting \eqref{eq:action} with a unit coefficient for the Einstein term gives
\begin{equation}
 I=\frac1{2\kappa_5^2}\int\dd^5x\sqrt{-g}
 \left[R-\frac12(\partial\varphi)^2-4V(\varphi/2)\right],
 \qquad\varphi=2p.
 \label{eq:canonical}
\end{equation}
For one scalar at zero density and fixed source,
\begin{equation}
 \frac\zeta\eta=\left(\frac{\dd\varphi_h}{\dd\log\ent}\right)^2,
 \qquad
 Z\equiv\frac\zeta\ent=\frac1\pi
       \left(\frac{\dd\ph}{\dd\log\ent}\right)^2.
 \label{eq:EO}
\end{equation}
Using $p$ as if it were $\varphi$ would miss a factor of four in the viscosity. This normalization agrees with the result for the PW effective action in \cite{BuchelEO}, expressed here in terms of $p$.

The high-temperature scalar asymptotics select the positive sign of $\dd\ph/\dd\log\ent$. We continue this sign along the monotonic thermal branch considered here, obtaining
\begin{equation}
 \ph(\ent)=\ph(\ent_\star)+
 \int_{\log\ent_\star}^{\log\ent}\sqrt{\pi Z(u)}\,\dd u.
 \label{eq:horizon-integral}
\end{equation}
The viscosity alone fixes neither this sign nor the integration constant. The high-temperature scalar solution supplies
\begin{equation}
 \ph=-\frac\pi4z+\cO(z^2)=-\frac1{4\pi T^2}+\cO(T^{-4}),
 \qquad Z=\frac\pi{36}z^2+\cO(z^3).
 \label{eq:EOanchor}
\end{equation}
The high-temperature boundary condition, $\lim_{T\to\infty}\ph(T)=0$ at fixed mass, fixes the integration constant. Here the limit refers to temperature, not radial position. Setting $\ph=0$ at the highest finite temperature would be incorrect. We integrate a controlled asymptotic tail to fix its nonzero value there and then use \eqref{eq:horizon-integral}. The entropy coordinate avoids numerical differentiation of the viscosity or an independent estimate of the speed of sound. It also makes the correlated nature of the resulting endpoint uncertainty explicit.

\subsection{Prepared input and the controlled B2 experiment}
The bulk-viscosity data were verified using both the Eling--Oz
horizon formula and an independent fluctuation calculation
\cite{ElingOz,BuchelCremoniniEarly}. We use 406 matched thermodynamic
and bulk-viscosity data points to reconstruct $\ph(T)$ through
\eqref{eq:horizon-integral}, with the integration constant fixed by
the high-temperature asymptotics. Varying the high-temperature fit
and interpolation changes the reconstructed horizon scalar by less
than $0.003\%$ over the training interval.

Experiment B2 supplies this reconstructed $\ph(T)$ to the networks,
retaining the architecture, initialization, training budget,
collocation points and loss weights of B1. Replace the free horizon output by
\begin{equation}
 J=K^2\ph^{\rm transport}(\ent),\qquad
 N_p(r,T)\longrightarrow N_p(r,T)-N_p(1,T)+J
 \label{eq:B2}
\end{equation}
in the scalar ansatz. The endpoint value is fixed, while the radial shape and horizon derivative remain adjustable. This introduces no new loss weight. Agreement with the imposed horizon scalar is now an input check, not an independent prediction. The remaining slopes still depend on $V$ and $V'$ through \eqref{eq:inverse-horizon}; the bulk interpolation, gauge coefficient and potential must still be learned.

High-temperature data are used only to fix the integration constant in the reconstruction of $\ph(T)$. The networks are trained over the same moderate-temperature interval as in B1, so the comparison isolates the effect of adding bulk-viscosity information.

\subsection{Reconstruction with transport input}
B2 was trained for 200,000 updates, with checkpoint selection based on equation residuals. Neither the exact potential nor the forward comparison errors entered that selection. Table~\ref{tab:transport-potential} compares B1 and B2 on the same sampled scalar grid, restricted to their common coverage. The maximum relative error in $U$ falls by a factor of about 13.5, as illustrated by the green B2 and orange B1 curves in figure~\ref{fig:potential-reconstruction}(b). The maximum absolute error in $V'$ falls by about 4.2 (table~\ref{tab:transport-potential}). The horizon trajectory has supplied useful information about the interaction that was only weakly constrained by thermodynamics.
\begin{table}[tb]
\centering
\begin{tabular}{lrr}
\toprule
Inverse diagnostic & B1: fixed $v_3$ & B2: with transport\\
\midrule
Maximum relative $U$ error (\%) & 0.390(7) & 0.0290(3)\\
Maximum relative $U$ error, $|p|\geq0.02$ (\%) & 0.353(5) & 0.0290(3)\\
Maximum absolute $V'$ error & $1.57(6)\times10^{-4}$ & $3.73(6)\times10^{-5}$\\
Free $v_4$ & $-2.24(2)$ & $-0.340(7)$\\
Common 193-node audit loss & $1.74(4)\times10^{-6}$ & $1.44(1)\times10^{-6}$\\
\bottomrule
\end{tabular}
\caption{Effect of the transport-derived horizon trajectory. Potential errors are evaluated on the common sampled scalar interval, approximately $[-0.1432,0]$. The audit loss is evaluated at the training temperatures. The cubic coefficient is supplied in both experiments; the horizon scalar is also supplied in B2 and is therefore not listed as a recovered observable.}
\label{tab:transport-potential}
\end{table}

The free quartic coefficient becomes $v_4=-0.340(7)$, within $2.20(5)\%$ of $-1/3$, without a quartic prior. This is a useful diagnostic rather than a new optimization target. Although the thermodynamic extraction in appendix~\ref{app:highT-recovery} determines $v_4$, we leave it unconstrained in the inverse experiments to test what bulk-viscosity information adds. B2 uses this additional observable over the thermal branch. The improvement is obtained at moderate temperature without training the geometry network on the high-temperature tail.

\subsection{Independent forward thermodynamics and viscosity}
We freeze the learned potential and independently solve the bulk equations at the same 27 temperatures, without supplying the reference entropy or horizon scalar. The resulting thermal family determines the bulk viscosity through
\begin{equation}
 \frac{\zeta}{\ent}=\frac1\pi\left[
 \frac{\dd\ph/\dd\log T}{\dd\log\ent/\dd\log T}\right]^2.
 \label{eq:Zforward}
\end{equation}
We check numerical convergence by varying the endpoint cutoffs and the temperature differentiation procedure.

Table~\ref{tab:transport-forward} shows the independent forward comparison. Thermodynamics remains very accurate and improves in every listed maximum-error diagnostic. More significantly, the maximum viscosity error falls by about a factor of four, both at the training temperatures and at the additional cold temperatures. The largest remaining viscosity discrepancy occurs at the coldest state in each subset.
\begin{table}[tb]
\centering
\begin{tabular}{lrr}
\toprule
Forward observable or subset & B1 & B2\\
\midrule
$\ent$, all 27 temperatures & 0.000384(9) & 0.000155(8)\\
$\press$, all 27 temperatures & 0.0000661(1) & 0.0000447(9)\\
$\energy$, all 27 temperatures & 0.000527(0) & 0.000219(5)\\
$\energy-3\press$, all 27 temperatures & 0.0122(3) & 0.00546(3)\\
$R-1$, all 27 temperatures & 0.0222(6) & 0.00901(3)\\
$(\energy-3\press)/(T\ent)$, all 27 temperatures & 0.0118(4) & 0.00530(7)\\
\midrule
$\zeta/\ent$, 17 matched training temperatures & 0.360(5) & 0.0895(5)\\
$\zeta/\ent$, ten additional cold temperatures & 0.295(1) & 0.0752(4)\\
\bottomrule
\end{tabular}
\caption{Maximum relative forward errors in percent, using only the frozen potential, temperature and source. The first viscosity row compares directly with the fluctuation data. The ten additional temperatures have no exactly matching fluctuation-data rows, so the second compares with independent PW forward solutions at those same temperatures. Thermodynamic comparisons use the conventional reference solutions in both subsets.}
\label{tab:transport-forward}
\end{table}

\begin{figure}[p]
 \centering
 \includegraphics[width=.92\textwidth]{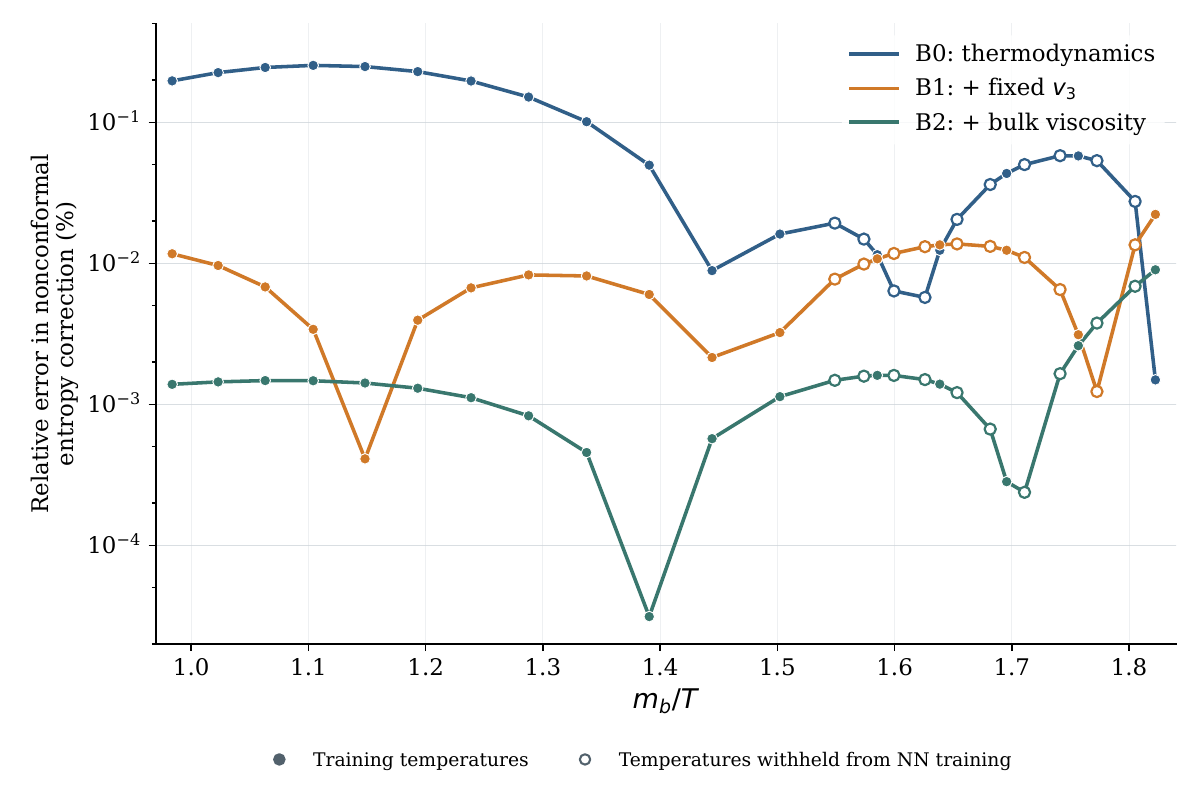}
 \caption{Recovery of the nonconformal equation of state in B0, B1
 and B2. The vertical axis shows
 $100\left|(R_{\rm reconstructed}-1)/(R_{\rm PW}-1)-1\right|$,
 where $R=\ent/(\pi^4T^3)$ as in \eqref{eq:zR}.
 Each curve comes from independently solving the bulk equations with
 the corresponding frozen learned potential. Filled circles mark
 training temperatures; open circles mark temperatures withheld from
 NN training. Lines connect sampled points. In B2, the full transport
 curve is used to construct the horizon constraint.}
 \label{fig:eos-reconstruction}

\medskip

 \centering
 \includegraphics[width=.92\textwidth]{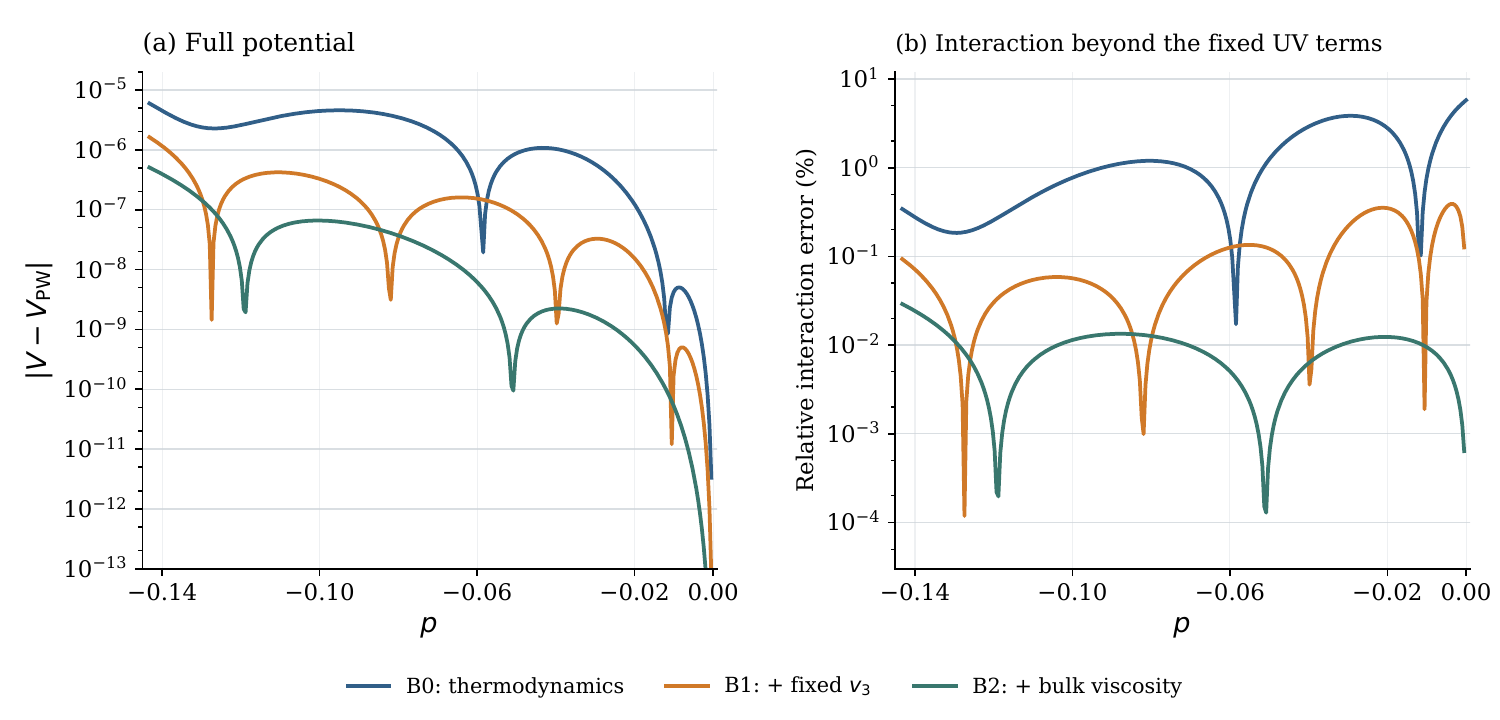}
 \caption{Recovery of the PW potential in B0 (blue), B1 (orange)
 and B2 (green). Panel (a) shows the absolute potential error
 $|V-V_{\rm PW}|$. Panel (b) shows the relative interaction error
 $100|U/U_{\rm PW}-1|$, with $U=(V+3+2p^2)/p^3$ as in
 \eqref{eq:U}. The curves use samples within the common scalar
 interval, approximately $-0.1432\leq p\leq0$.
 The finite limit $U(0)=v_3$ is used at the boundary; exact zero
 errors are omitted from the logarithmic axes.}
 \label{fig:potential-reconstruction}
\end{figure}

Figure~\ref{fig:eos-reconstruction} compares the nonconformal entropy
correction predicted by the three reconstructed potentials with PW.
Fixing the cubic interaction reduces the maximum relative error from
$0.254\%$ (blue B0) to $0.0223\%$ (orange B1); incorporating bulk
viscosity reduces it further to $0.00901\%$ (green B2), an overall
improvement of about a factor of 28.
The improvement extends across most of the temperature interval and
is also visible at temperatures withheld from NN training, although
it is not pointwise monotonic.

The potential comparison in figure~\ref{fig:potential-reconstruction}
shows a larger improvement in the recovered interaction. Its maximum
relative error decreases from $5.768\%$ in B0 to $0.3907\%$ in B1
and $0.02903\%$ in B2, an overall reduction by about a factor of 199.
Thus the added physics improves both the equation of state and the
potential, with a particularly strong effect on the interaction that
thermodynamics alone constrains weakly.

{\emergencystretch=1em
The largest relative change between three- and five-point viscosity estimates is $0.0143(9)\%$ for PW, $0.0148(9)\%$ for B1 and $0.0145(0)\%$ for B2. The largest residual in the radially conserved enthalpy identity is approximately $4.43\times10^{-9}$. At the 17 directly matched temperatures, the independent PW viscosity agrees with the fluctuation data within $1.88(8)\times10^{-5}\%$.\par}

Using the independent PW solver as a common reference at all 27 temperatures gives maximum viscosity errors $0.360(5)\%$ and $0.0895(6)\%$ for B1 and B2, respectively, with RMS errors $0.109(2)\%$ and $0.0280(5)\%$. The reduction is larger than the measured derivative sensitivity. These checks establish that the improved potential supports its own more accurate thermal and transport response, rather than merely reproducing endpoint values inside the inverse ansatz.

\FloatBarrier
\section{Discussion and extensions}
\label{sec:discussion}
The completed experiments separate three questions that are easily conflated. Can a neural representation solve the bulk equations? Can a learned potential reproduce thermodynamics in an independent calculation? How well is the interaction itself recovered? The linear benchmark and nonlinear direct runs answer the first question at useful accuracy on the tested intervals. The frozen-potential forward solves answer the second. The discrepancy between the interaction coefficients and the forward EoS shows why the third question needs additional diagnostics and, potentially, additional observables.

Our implementation incorporates five additional physics inputs relative
to Bea et al.~\cite{Bea2024}:
\begin{enumerate}
 \item The thermodynamic identity $\energy+\press=T\ent$
 fixes $A_{40}$ through \eqref{eq:A40}.
 \item The trace Ward identity fixes $\Puv$ from the energy density
 and pressure through \eqref{eq:ward-P20}.
 \item Horizon regularity imposes the relations \eqref{eq:horizon}
 exactly in the network construction.
 \item The first law along the thermal family supplies the
 differential constraint \eqref{eq:firstlaw}, included in the
 loss \eqref{eq:directloss}.
 \item Bulk viscosity, together with the high-temperature scalar
 asymptotics, fixes $\ph(T)$ through the integrated horizon relation
 \eqref{eq:horizon-integral}.
\end{enumerate}

Known physics is most useful when it changes which information the network has to infer. Residual normalization prevents boundary coefficients from becoming cheap errors. The Ward identity fixes ultraviolet one-point data once the thermodynamic scheme is chosen. Horizon regularity relates slopes to the potential. High-temperature thermodynamics supplies a cubic constraint before moderate-temperature training. None of these ingredients replaces optimization, but each removes an ambiguity or an avoidable conditioning problem. Stronger subtraction alone did not improve the linear experiment, whereas targeted radial sampling improved the nonlinear residuals. These observations argue for diagnosing the location and meaning of errors before changing capacity or pursuing a smaller aggregate loss.

The high-temperature EOS analysis in appendix~\ref{app:highT-recovery} recovered $v_4$ to $0.37\%$, showing that a quartic prior can be obtained from sufficiently precise thermodynamics alone. We deliberately leave $v_4$ unconstrained in the inverse experiments to test what bulk-viscosity information adds. The integrated bulk-viscosity relation constrains a function over the thermal branch. By contrast, extracting $v_4$ from high-temperature thermodynamics requires resolving a strongly suppressed contribution. With the source fixed to $m=1$, the scalar behaves as $p=\cO(T^{-2})$, so $v_4p^4=\cO(T^{-8})$, compared with the quadratic term of order $T^{-4}$. For the PW value $v_4=-1/3$, the ratio of the quartic and quadratic contributions is $p^2/6$, approximately $1.88\times10^{-4}$, or $0.0188\%$, at the horizon of the hottest state used in training. In the completed experiment this input reduces the maximum sampled interaction error by about thirteenfold (orange B1 to green B2 in figure~\ref{fig:potential-reconstruction}(b)) and the independent forward viscosity error by about fourfold (table~\ref{tab:transport-forward}), while also improving the forward EoS (figure~\ref{fig:eos-reconstruction}). The free quartic coefficient $v_4$ moves close to its PW value as an outcome, without being supplied as a constraint. Thermodynamics and transport constrain complementary information about the bulk theory on the tested interval.

There are limitations. Most controlled comparisons here use one seed, and the sampling comparisons are continuations of one baseline. Ensembles are needed to distinguish repeatable information gain from optimization variability. All potential statements are restricted to the scalar interval explored by the black branes. No result constrains the potential at larger fields. The fixed-cubic experiment uses an exact prior; a subsequent test should propagate the uncertainty of the thermodynamic cubic fit. The numerical EoS is far more precise than most phenomenological inputs. With uncertain data, hard thermodynamic endpoint values should be replaced by a consistent smooth EoS representation and uncertainty-aware constraints. The first law and Ward identity should organize that representation rather than force incompatible noisy inputs to agree independently.

The supergravity origin of PW suggests another possible prior, a superpotential \cite{FakeSupergravity}. Its mere local existence for a single scalar generally supplies little independent restriction: representing a potential through a superpotential amounts locally to a first-order nonlinear equation with a choice of branch. Global regularity, a specified ultraviolet branch and an admissible infrared completion can be restrictive, but those are additional physical assumptions. We have not imposed such an input. In particular, a phenomenological equation of state need not describe a supersymmetric theory or extend to zero temperature: it may instead have a finite-temperature endpoint (as in the $\Ntwo$ model discussed here).

\subsection{Testing the assumed operator content}
Several active scalars and deformation scales provide a natural next test. The number of fields, their dimensions, their sources and their symmetry restrictions are part of the hypothesis supplied to the inverse problem. For a scalar, the assumed dimension fixes $m_I^2L_{\rm AdS}^2=\Delta_I(\Delta_I-4)$. High-temperature thermodynamics can reveal the dominant deformation, but finite-precision data over a finite temperature interval need not reveal every subleading operator. A successful one-scalar EoS reconstruction is therefore conditional on its assumed field content.

A controlled experiment would generate thermodynamics with both a dimension-three and a dimension-two deformation, then deliberately reconstruct those data with a single scalar whose dimension is fixed to three. At sufficiently high temperature the dimension-three contribution dominates, so the hypothesis cannot be rejected merely from the leading nonconformal power. The one-field potential must instead absorb the missing dimension-two sector into effective nonlinear interactions. Varying the second source would show how long such a description remains viable, how its potential is distorted, and whether withheld transport exposes the missing field before the EoS fit deteriorates. The two-scalar PW model offers a concrete setting for this test. Transport can falsify an insufficient model without necessarily determining a unique microscopic operator content. With several scalars, its horizon relation constrains a combination of their motion through field space, rather than fixing every horizon component separately. We hope to return to these questions in the future.

\subsection{Learning the fluctuation sector and testing the transport input}
The completed B2 experiment incorporates bulk viscosity through the integrated horizon relation. A complementary implementation would train the properly ultraviolet-normalized, zero-frequency gauge-invariant scalar fluctuation together with the background and potential \cite{BuchelCremoniniEarly}. We use the fluctuation normalization for the canonical scalar $\varphi=2p$, following \cite{BuchelCremoniniEarly}. Writing its horizon value as $z_h$, the measured input imposes
\begin{equation}
 z_h(T)^2=9\pi\,\frac{\zeta(T)}{\ent(T)}.
 \label{eq:future-fluctuation-input}
\end{equation}
The fluctuation equation contains the second derivative of the potential. Horizon regularity fixes a relation between the fluctuation and its radial derivative at the horizon that involves this local curvature, while the ultraviolet normalization fixes the amplitude of the homogeneous solution. Fixing only the horizon logarithmic derivative would not fix $z_h$: it is a functional of the full background and potential. A fluctuation network would make that dependence explicit and provide a separate implementation of the transport constraint. In the present one-scalar, two-derivative setting, the fluctuation and Eling--Oz descriptions encode the same viscosity observable; using both is a consistency and conditioning test, not two independent measurements. For a constant Gauss--Bonnet coupling, the fluctuation calculation agrees with the entropy-normalized Eling--Oz relation \cite{BuchelGB,DemircikAnisotropic}. Whether this agreement extends to a scalar-dependent Gauss--Bonnet coupling remains unknown and requires a separate calculation.

An idealized control would replace the transport-derived horizon trajectory by the exact numerical PW trajectory while withholding the PW potential. Matching the remaining priors, state range and optimization budget to B2 would separate uncertainty in the transport integration and high-temperature anchor from the difficulty of learning the geometry and potential. A second version, keeping only $V=-3-2p^2+\cO(p^3)$ and leaving the cubic free, would isolate how much ultraviolet interaction information a perfect horizon trajectory supplies by itself. These controls are diagnostic uses of otherwise unavailable geometry, not proposed inputs for phenomenological reconstruction.

The high-temperature transport expansion can also be developed further using the higher-order perturbative PW background. Its role would be to check the finite-temperature fluctuation calculation and the asymptotic integration constant, and to quantify the sensitivity to a finite upper-temperature cutoff. The present transport data have already been checked by both the horizon and fluctuation approaches; a higher-order analytic anchor would be an additional controlled check. These calculations should assess the information and uncertainty of the input, rather than pursue another small digit without a reconstruction question.

\subsection{Higher derivatives and successive transport constraints}
The horizon values of the potential and its derivatives have a physical role beyond enforcing regularity. For example, let $\mathcal U(\varphi)$ denote the scalar potential in an action written as $R+12-\tfrac12(\partial\varphi)^2-\mathcal U(\varphi)$, so that $\varphi=2p$ and $\mathcal U(2p)=12+4V(p)$ in our conventions. For the constant Gauss--Bonnet coupling normalized as in \cite{BuchelGB},
\begin{equation}
 \frac{\eta}{\ent}=\frac1{4\pi}
 \left[1+\frac{\lambda_{\rm GB}}3(\mathcal U_h-12)\right]
 =\frac1{4\pi}\left[1+\frac{4\lambda_{\rm GB}}3V(p_h)\right].
 \label{eq:future-GB-shear}
\end{equation}
Thus a horizon potential value entering the background constraints also controls shear transport once Einstein universality is lifted. Potential derivatives can likewise enter fluctuation equations and more detailed hydrodynamic response. An identification of a particular derivative with a particular higher-order transport coefficient requires an explicit calculation; we do not assume, for example, that $V''_h$ alone determines a relaxation time.

This suggests a further controlled sequence: reconstruct from thermodynamics, then from thermodynamics plus shear viscosity, and finally from thermodynamics plus both shear and bulk viscosities. At each stage observables withheld from training test the reconstructed theory. The frameworks of \cite{BuchelCremoniniEarly,BuchelGB,BuchelCharged} provide the relevant transport relations, including charged higher-derivative models. General higher-derivative interactions require the appropriate corrected equations and entropy prescription; the two-derivative formulas cannot simply be carried over unchanged. Such tests could distinguish insufficient matter content from additional gravitational interactions that equilibrium data constrain weakly.

\subsection{A bulk $U(1)$ sector: the magnetized gauged PW benchmark}
A particularly useful intermediate problem is the gauged $\Ntwo$ model studied in \cite{BuchelUmbertMagnetic}. Its PW scalar sector is supplemented by a bulk $U(1)$ gauge field, and its thermodynamics has been computed in an external magnetic field. This supplies a known top-down benchmark with both scalar and gauge sectors before attempting a charged reconstruction from phenomenological data. The magnetic field also makes the stress tensor anisotropic, so the geometry, thermodynamic dictionary and loss constraints must be enlarged consistently.

The magnetized setup is distinct from a fluid at finite charge density. A subsequent finite-density problem must treat the conserved $U(1)$ charge and its chemical potential explicitly, together with the associated thermodynamic identities and gauge-field regularity. This is a concrete step towards the hypothetical water question posed in the introduction, for which a conserved particle number cannot be omitted. Testing the enlarged architecture first on a known gauged supergravity model would separate implementation difficulties from uncertainty about the inferred gauge and scalar couplings.

These extensions can be combined with multibranch thermodynamics and with the uncertainty tests described above. Applying the programme to QCD-motivated thermodynamics and transport is the subject of ongoing work \cite{QCDprogress}. The purpose of the present first step is to establish a controlled language for deciding what has actually been reconstructed before enlarging the theory and its inputs.

\acknowledgments
I would like to thank Sera Cremonini, Pavel Kovtun and David Mateos for useful discussions. I am grateful to the Centro de Ciencias de Benasque Pedro Pascual for the inspiring environment in which this project began. This work was supported by the Natural Sciences and Engineering Research Council of Canada (NSERC).

\appendix
\section{Critical-temperature estimates from the high-temperature equation of state}
\label{app:critical-extrapolation}
\subsection{Numerical reference and normalization}
The numerical EOS consists of 356 black-brane solutions with
$10^{-4}\leq\rho_{11}\leq0.0356$ in steps of $10^{-4}$, using the
boundary parameter $\rho_{11}$ of \cite{BuchelThermo}.
The sampled range is $0.1294\lesssim m_b/T\lesssim2.324$ and includes
a short returning branch beyond the minimum temperature.
The PINN training interval is approximately
$0.984\lesssim m_b/T\lesssim1.822$; its cold endpoint has
$T/T_c\simeq1.275$. Across this interval the entropy deficit relative
to the conformal value ranges from $0.0920\%$ to $1.697\%$.

In the radial convention of \cite{BuchelThermo}, the conversions are
\begin{equation}
 u=\frac{\pi}{\sqrt2}e^{6a_h}\rho_{11},\qquad
 \frac{\ent}{N^2T^3}=\frac{\pi^2}{2}e^{12a_h},\qquad
 R\equiv\frac{2\ent}{\pi^2N^2T^3}=e^{12a_h}.
 \label{eq:archive-eos-map}
\end{equation}
This $R$ equals the normalized entropy in \eqref{eq:zR};
$u=\pi z/(4\sqrt6)$ in the unit-source convention.
Thus restoring $N^2$ changes the plotted normalization, not the EOS.
The first row has $u=0.0002221$ and the largest sampled value is
$u=0.07163$, at $\rho_{11}=0.0351$.
This gives $T_c/m_b\simeq0.4303$. Local polynomial interpolations
of degrees two through four over five and seven neighbouring rows
give the same displayed value. We use the critical point of this
more precise numerical dataset as the reference throughout the
comparison below; \cite{BuchelCritical} reported $T_c/m_b\simeq0.4299$.

\subsection{Truncation order and the singularity assumption}
Write the PW high-temperature series in the positive variable
\eqref{eq:u-critical} as
\begin{equation}
 R_N(u)=1-\sum_{n=2}^{N}b_nu^n,
 \qquad
 (b_2,b_3,b_4,b_5,b_6)
 \simeq(4.863,48.78,477.6,4.910\times10^3,5.283\times10^4).
 \label{eq:critical-RN}
\end{equation}
The first coefficient is exactly $48/\pi^2$.
The coefficients through fourth order follow from
appendix~\ref{app:highT} with the PW couplings.
The fifth- and sixth-order coefficients follow from the continuation
of the perturbative calculation with the known PW potential.
All calculations use the unrounded coefficients.
Figure~\ref{fig:PWentropyHT} shows $N=2,4,6$, corresponding to terms
through $(m_b/T)^4,(m_b/T)^8,(m_b/T)^{12}$, respectively.

A finite polynomial for $R$ has no finite-temperature branch point.
Indeed, at fixed mass, with $c_V=(\partial\energy/\partial T)_{m_b}$,
\begin{equation}
 \frac{c_V}{\ent}=3-2u\frac{R'(u)}{R(u)},\qquad
 c_s^2=\frac{R(u)}{3R(u)-2uR'(u)}.
 \label{eq:critical-cs}
\end{equation}
For polynomial $R$ and positive finite entropy the specific heat
cannot diverge. A zero of the polynomial entropy is not the desired
critical point. As a deliberately different prescription, one can
Taylor-expand the second expression in \eqref{eq:critical-cs}
through $u^N$ and find its first positive zero.
Table~\ref{tab:critical-orders} shows that these formal zeros lie
well outside the controlled high-temperature region and substantially
underestimate $T_c$. They are not physical singularities of $R_N$.

More information can be extracted if the leading singularity is
\emph{assumed} to be the positive-real square-root branch point
associated with the known fold. Its large-order consequence is
$b_n\propto u_c^{-n}n^{-3/2}$, up to subleading corrections.
Two finite-order radius estimates are then
\begin{equation}
 u_c^{\rm raw}(n)=\left|\frac{b_{n-1}}{b_n}\right|,\qquad
 u_c^{\rm sqr}(n)=\left(\frac{n-1}{n}\right)^{3/2}
                 \left|\frac{b_{n-1}}{b_n}\right|,
 \qquad \frac{T_c^{\rm est}}{m_b}=\frac1{\sqrt{24\pi u_c^{\rm est}}}.
 \label{eq:critical-ratio}
\end{equation}
The overall proportionality constant cancels in these ratios. No numerical critical location is inserted
in these estimates, but the exponent and the assumption that this
singularity dominates are additional inputs.
\begin{table}[tb]
\centering
\begin{tabular}{rrrrrr}
\toprule
$N$ & $c_s^2$ zero: $T_c/m_b$ & Raw ratio & Square-root ratio
& $u_c^{\rm sqr}$ & Square-root error (\%)\\
\midrule
2 & 0.1838 & --- & --- & --- & ---\\
3 & 0.2594 & 0.3647 & 0.4944 & 0.05427 & $+14.89$\\
4 & 0.3021 & 0.3603 & 0.4471 & 0.06635 & $+3.902$\\
5 & 0.3290 & 0.3693 & 0.4365 & 0.06960 & $+1.446$\\
6 & 0.3473 & 0.3778 & 0.4331 & 0.07070 & $+0.6538$\\
\bottomrule
\end{tabular}
\caption{Critical-temperature diagnostics from successive orders of the PW
high-temperature expansion. The current numerical reference is
$T_c/m_b\simeq0.4303$; errors use its unrounded
value. Raw and square-root columns both report $T_c/m_b$.
The square-root estimate at sixth order is encouraging, but its
$0.65\%$ agreement is conditional on the assumed singularity and
is not an uncertainty estimate.}
\label{tab:critical-orders}
\end{table}
The raw sixth-order estimate is $12.21\%$ low, whereas the zero of the
sixth-order sound-speed series is $19.29\%$ low.
The progressively improving square-root estimates therefore carry
useful evidence about the known PW series, but do not establish
sub-percent prediction of a critical temperature from an arbitrary
finite-order EOS. The available terms through sixth order cannot
exclude other singularities or bound the omitted large-order corrections.

Extracting the fifth- and sixth-order coefficients directly from the numerical EOS does not yield a stable critical-temperature estimate: the result depends strongly on the fitting interval and truncation order. The successful estimate above therefore relies on the perturbatively calculated coefficients. This sensitivity does not invalidate the extraction of the lower-order cubic and quartic interactions described in appendix~\ref{app:highT}.

\FloatBarrier
\section{High-temperature extraction of the scalar potential}
\label{app:highT}
We develop the high-temperature expansion in Eddington--Finkelstein coordinates, extending the perturbative results of \cite{BenincasaBuchelStarinets} to the orders needed to extract the cubic and quartic coefficients of the scalar potential.

\subsection{High-temperature expansion of the normalized entropy density}
\label{app:highT-expansion}
Our goal is to relate the high-temperature equation of state to the coefficients $v_3$ and $v_4$ of the scalar potential,
\begin{equation}
 V(p)=-3-2p^2+v_3p^3+v_4p^4+\cdots.
\end{equation}
Recall that
\begin{equation}
 R\equiv\frac{\ent}{\pi^4T^3}
\end{equation}
is the entropy density divided by its conformal value at the same temperature. Thus $R=1$ in the conformal limit, and $R-1$ measures the nonconformal correction.

For this calculation, we place the horizon at $x=1$ and choose the radial gauge in which the undeformed metric functions are $s_0=1$ and $a_0=1-x^4$. The expansion parameter $q$ is the scalar-source coefficient:
\begin{equation}
 p(x)=q\,x^2\log x+\cdots\qquad (x\to0).
\end{equation}
Small $q$ describes a small mass deformation relative to the temperature.

The required expansions are
\begin{align}
 p={}&qp_1+q^2v_3p_2+q^3(p_{3,0}+v_3^2p_{3,a}+v_4p_{3,b})+\cO(q^4),\nonumber\\
 s={}&1+q^2s_2+q^3v_3s_3+q^4(s_{4,0}+v_3^2s_{4,a}+v_4s_{4,b})+\cO(q^5),\nonumber\\
 a={}&1-x^4+q^2a_2+q^3v_3a_3+q^4(a_{4,0}+v_4a_{4,a}+v_3^2a_{4,b})+\cO(q^5).
 \label{eq:perturbansatz}
\end{align}
All coefficient functions depend on $x$. Only $p_1$ carries the prescribed logarithmic source; the higher scalar corrections have no additional source. We impose the boundary metric normalization and regularity at the horizon at each order. The scalar is needed through order $q^3$, and the metric through order $q^4$.

The horizon geometry determines the temperature and entropy density:
\begin{equation}
 \frac{T_{\rm canonical}}{T_0}=-\frac{a'(1)}4,
 \qquad \frac{\ent_{\rm canonical}}{\ent_0}=s(1)^3,
 \qquad T_0=\frac1\pi,\quad \ent_0=\pi.
\end{equation}
Here the prime denotes an $x$ derivative. Consequently,
\begin{equation}
 R=\frac{s(1)^3}{[-a'(1)/4]^3}.
\end{equation}
Expanding this expression gives $R$ as a series in $q$.

To compare with the physical equation of state, we must express the result in terms of the mass-to-temperature ratio. The appropriate dimensionless variable is
\begin{equation}
 z=\frac{q}{\pi^2T_{\rm canonical}^2}
  =\frac1{\pi^2T^2}
\end{equation}
in the fixed-source convention $m=1$, where $T$ is measured in source units. Although $z=q$ at leading order, the temperature corrections must be included to obtain the quartic coefficient correctly.

After this conversion, the normalized entropy density is
\begin{equation}
 R(z)=1-\frac12z^2+\frac23c_2v_3z^3
 +(K_{40}+c_{3a}v_3^2+c_{3b}v_4)z^4+\cO(z^5),
\end{equation}
as quoted in \eqref{eq:Reos}. Here $c_2,c_{3a},c_{3b}$ are the coefficients of the nonlogarithmic $x^2$ terms in $p_2,p_{3,a},p_{3,b}$, respectively. The constant $K_{40}$ is the contribution at order $z^4$ independent of $v_3$ and $v_4$. Their numerical values are listed in table~\ref{tab:highT-constants} and follow from the generic perturbative bulk equations, without assuming the PW interaction coefficients.

\begin{table}[htbp]
\centering
\begin{tabular}{cccc}
\toprule
$c_2$ & $c_{3a}$ & $c_{3b}$ & $K_{40}$\\
\midrule
$-4.43(1)$ & $-17.7(8)$ & $1.44(4)$ & $0.701(4)$\\
\bottomrule
\end{tabular}
\caption{Numerical coefficients in the generic high-temperature
entropy expansion \eqref{eq:Reos}. The displayed values are rounded; the extraction in appendix~\ref{app:highT-recovery} uses the unrounded coefficients.}
\label{tab:highT-constants}
\end{table}

The quadratic term is universal for the fixed scalar mass and normalization. The cubic term determines $v_3$; once $v_3$ is known, the quartic term determines $v_4$. The next subsection applies these relations to numerical thermodynamic data.

\subsection{Thermodynamics-only recovery of the cubic and quartic interactions}
\label{app:highT-recovery}
In temperature units, $T=1$, the small parameter is the dimensionless
mass deformation. Define
\begin{equation}
 \delta_1=-u=-\frac1{24\pi}\left(\frac{m_b}{T}\right)^2,
 \qquad z=-\frac{4\sqrt6}{\pi}\delta_1.
\end{equation}
Equation~\eqref{eq:Reos} then reads
\begin{align}
 R={}&1-\frac{48}{\pi^2}\delta_1^2
 -\frac{256\sqrt6}{\pi^3}c_2v_3\delta_1^3\nonumber\\
 &+\frac{9216}{\pi^4}
 (K_{40}+c_{3a}v_3^2+c_{3b}v_4)\delta_1^4
 +\cO(\delta_1^5).
 \label{eq:temperature-unit-eos}
\end{align}
The constants in table~\ref{tab:highT-constants} are determined by the
generic perturbative bulk equations. Their values do not assume the
PW cubic or quartic coupling.

Using the numerical EOS with $|\delta_1|\leq0.01$, we fit
\begin{equation}
 R_{\rm fit}=1+\sum_{n=2}^{7}\widehat c_n\delta_1^n.
 \label{eq:temperature-unit-fit}
\end{equation}
All six coefficients are fitted, including the quadratic coefficient
as a normalization check. The fifth-, sixth- and seventh-order terms
absorb higher-order contamination. The first three fitted coefficients
are approximately $-4.863$, $48.78$ and $-477.7$; their relative
differences from the perturbative PW values are $4.9\times10^{-8}$,
$9.1\times10^{-6}$ and $3.4\times10^{-4}$, respectively.
The comparison uses the unrounded coefficients. Truncating the fit
at fourth order gives a less stable quartic estimate.

The inferred potential coefficients follow algebraically:
\begin{align}
 v_3^{\rm EOS}&=-\frac{\pi^3\widehat c_3}{256\sqrt6\,c_2},\nonumber\\
 v_4^{\rm EOS}&=\frac{\pi^4\widehat c_4/9216-K_{40}
                         -c_{3a}(v_3^{\rm EOS})^2}{c_{3b}}.
 \label{eq:temperature-unit-inversion}
\end{align}
The fit gives
\begin{equation}
 v_3^{\rm EOS}\simeq0.544326,\qquad
 v_4^{\rm EOS}\simeq-0.3346,
 \label{eq:temperature-unit-couplings}
\end{equation}
corresponding to relative errors of $0.00091\%$ and $0.37\%$
against \eqref{eq:Vuv}. Thus both interactions can be recovered
directly from sufficiently precise high-temperature thermodynamics,
without PINN training.

\subsection{Triangular differential equations}
We now give the perturbative bulk equations used to obtain the high-temperature entropy expansion in appendix~\ref{app:highT-expansion}. They are solved successively: the leading scalar profile determines the first metric corrections, which then enter the higher-order scalar equations. Boundary normalization and horizon regularity fix the solutions at each order, yielding the coefficients needed to extract $v_3$ and $v_4$ from thermodynamics.

Primes in this subsection mean $x$ derivatives. Define
\begin{equation}
 \mathscr L f=f''+\frac{x^4+3}{x(x^4-1)}f'-\frac4{x^2(x^4-1)}f.
\end{equation}
The linear scalar and second-order sources obey
\begin{align}
 0&=\mathscr Lp_1,\nonumber\\
 0&=s_2''+\frac23(p_1')^2,\nonumber\\
 0&=a_2'-\frac4x a_2+(6-2x^4)s_2'
       +\frac23x(1-x^4)(p_1')^2+\frac8{3x}p_1^2,\nonumber\\
 0&=\mathscr Lp_2+\frac{3p_1^2}{x^2(x^4-1)}.
\end{align}
At the next order,
\begin{align}
 0&=s_3''+\frac43p_1'p_2',\nonumber\\
 0&=a_3'-\frac4x a_3+(6-2x^4)s_3'
 -\frac43x(x^4-1)p_1'p_2'-\frac4{3x}p_1(p_1^2-4p_2),\nonumber\\
 0&=\mathscr Lp_{3,a}+\frac{6p_1p_2}{x^2(x^4-1)},\qquad
 0=\mathscr Lp_{3,b}+\frac{4p_1^3}{x^2(x^4-1)},\nonumber\\
 0&=\mathscr Lp_{3,0}+\left[3s_2'-\frac{a_2'}{x^4-1}
 +\frac{4x^3a_2}{(x^4-1)^2}\right]p_1'
 -\frac{4p_1a_2}{x^2(x^4-1)^2}.
\end{align}
The entropy sector at fourth order is
\begin{align}
 0&=s_{4,a}''+\frac23(p_2')^2+\frac43p_1'p_{3,a}',\nonumber\\
 0&=s_{4,b}''+\frac43p_1'p_{3,b}',\nonumber\\
 0&=s_{4,0}''+\frac43p_1'p_{3,0}'+\frac23s_2(p_1')^2.
\end{align}
The remaining temperature equations are
\begin{align}
 0={}&a_{4,b}'-\frac4x a_{4,b}-2(x^4-3)s_{4,a}'
 -\frac23x(x^4-1)(p_2')^2\nonumber\\
 &-\frac4{3x}p_2(3p_1^2-2p_2)-\frac43x(x^4-1)p_{3,a}'p_1'
 +\frac{16}{3x}p_{3,a}p_1,\nonumber\\
 0={}&a_{4,a}'-\frac4x a_{4,a}-2(x^4-3)s_{4,b}'
 -\frac43x(x^4-1)p_{3,b}'p_1'-\frac4{3x}p_1^4+\frac{16}{3x}p_{3,b}p_1,\nonumber\\
 0={}&a_{4,0}'-\frac4x a_{4,0}+(6-2x^4)s_{4,0}'+4x(s_2')^2\nonumber\\
 &+\left[2a_2+\frac83p_1^2-\frac23x^6(p_1')^2
 +\frac23x^2(p_1')^2+2x^4s_2-6s_2\right]s_2'\nonumber\\
 &+\frac23xa_2(p_1')^2-\frac43x(x^4-1)p_{3,0}'p_1'
 +\frac{16}{3x}p_1p_{3,0}.
\end{align}
This triangular system separates the $1$, $v_3^2$ and $v_4$ structures without assuming their PW values. It also explains why an unrestricted finite-temperature polynomial must contain nuisance terms above quartic order when extracting the cubic coupling from data away from $z=0$.

\section{Linearized solution and finite ultraviolet residuals}
\label{app:linear}
For the analytic reduction, introduce the radial coordinate
\begin{equation}
 t=\frac{(1+k)r}{1+kr},\qquad
 \frac{\dd t}{\dd r}=\frac{1+k}{(1+kr)^2}.
 \label{eq:tmap}
\end{equation}
It maps the boundary and horizon to zero and one. The scalar equation becomes
\begin{equation}
 p_{tt}-\frac{3+t^4}{t(1-t^4)}p_t+\frac4{t^2(1-t^4)}p=0.
\end{equation}
Writing $p=t^2f(z)$ with $z=1-t^4$ gives
\begin{equation}
 z(1-z)f_{zz}+(1-2z)f_z-\frac14f=0.
\end{equation}
Horizon regularity selects the hypergeometric branch in \eqref{eq:exact}.
Using ${}_2F_1(\tfrac12,\tfrac12;1;1-t^4)
=\tfrac2\pi\log(4/t^2)+\cO(t^4\log t)$,
\begin{equation}
 p_{\rm ex}=\frac{r^2}{(1+kr)^2}\left[\log r+\log\frac{1+k}{2}-\log(1+kr)
 +\cO(t^4\log t)\right],
\end{equation}
which fixes the unit source and \eqref{eq:linear-exact-coeff}.

To expose the finite residuals in the single-state rescaling test,
write the remainders near the boundary in terms of two regular
functions $A(r)$ and $B(r)$:
\begin{align}
 p&=\frac{r^2\log r}{(1+kr)^2}+\Puv r^2+r^3A(r),\nonumber\\
 \nu_p&=\frac{r[2\log r+1+kr]}{(1+kr)^3}
              +2\Puv r+r^2B(r).
\end{align}
Substitution into \eqref{eq:linear-residual} gives
\begin{align}
 \frac{C_p}{r^2}={}&3A+rA'-B,\nonumber\\
 \frac{E_p}{r^3}={}&r(1+kr)aB'
 +\big[-(1+kr)a+r(1+kr)a'+3rak\big]B+4(1+kr)A\nonumber\\
 &+\left[2k(1+kr)(1+3kr)
 -\frac{2(1+k)^4r^3(2+3kr)}{(1+kr)^2}\right]\Puv\nonumber\\
 &+k-\frac{(1+k)^4r^3}{(1+kr)^5}
        \big[4\log r+(1+kr)(4+kr)\big].
\end{align}
The limiting $r^3\log r$ term is zero. At $r=0$ the equations give
\begin{equation}
 3A(0)-B(0)=0,\qquad4A(0)-B(0)+2k\Puv+k=0,
\end{equation}
and hence $A(0)=-2k\Puv-k$, precisely the next term in
\eqref{eq:UVp}. The rescaling makes this boundary information
contribute to the loss without suppression by powers of $r$.

\section{Nonlinear subtraction and boundary conditions}
\label{app:subtraction}
With $L=\log r$ and $Q=\Puv-\log w$, the logarithmic portions of the leading subtraction can be written
\begin{align}
 b_p&=\frac{r^2}{w^2}L+\beta\frac{r^4}{w^4}L(L+2Q-2),\nonumber\\
 b_s&=-\frac{r^4}{27w^3}L(6L+12Q-1),\qquad b_a=2wb_s.
\end{align}
The implemented bases \eqref{eq:bases} include the accompanying smooth terms. These rational expressions resum an infinite set of powers of $kr$ without pretending to remove every nonlinear logarithm. The coefficient $\beta$ depends on the cubic interaction.

For reference, the original radial scalar and transverse-metric equations are
\begin{equation}
 p''+\left(\frac{3s'}s+\frac{a'}a-\frac3r\right)p'
 -\frac{V'(p)}{r^2a}=0,\qquad
 s''+\frac23s(p')^2=0.
\end{equation}
The first-order constraint is equivalently \eqref{eq:Ea} with the consistency equations imposed. One diagnostic form of the remaining metric equation is
\begin{equation}
 a''+\left(\frac{3s'}s-\frac5r\right)a'
 +\left(\frac8{r^2}-\frac{6s'}{rs}\right)a
 +\frac{8V(p)}{3r^2}=0.
 \label{eq:audit-a}
\end{equation}
It is redundant for an exact solution but useful for testing an approximate neural solution. Define the first-order radial residual $E_s^{(r)}=\partial_r\nu_s+\frac23s\nu_p^2$. Off shell, transforming this residual gives
\begin{equation}
 E_s^{(r)}=\frac{K^2}{w^3}E_s+\frac{kK}{w^2}C_s,
\end{equation}
which explains why the preconditioned equation preserves the full solution set without being the same off-shell residual.

At the horizon the slope outputs obey \eqref{eq:inverse-horizon} in both the direct and inverse problems. In the transformed variables $S(1)=h/K$ and $A(1)=0$. Subtracting each base's horizon value and adding the corresponding fixed endpoint remainder enforces these field values while preserving a free network shape. Consistency with the independently parametrized slopes is still learned.

\bibliographystyle{JHEP}
\bibliography{references}

@book{QGPreview,
    author = "Casalderrey-Solana, Jorge and Liu, Hong and Mateos, David and Rajagopal, Krishna and Wiedemann, Urs Achim",
    title = "{Gauge/String Duality, Hot QCD and Heavy Ion Collisions}",
    eprint = "1101.0618",
    archivePrefix = "arXiv",
    primaryClass = "hep-th",
    reportNumber = "CERN-PH-TH-2010-316, MIT-CTP-4198, ICCUB-10-202",
    doi = "10.1017/9781009403504",
    isbn = "978-1-009-40350-4, 978-1-009-40349-8, 978-1-009-40352-8, 978-1-139-13674-7",
    publisher = "Cambridge University Press",
    year = "2014"
}

@article{CheslerYaffe,
    author = "Chesler, Paul M. and Yaffe, Laurence G.",
    title = "{Numerical solution of gravitational dynamics in asymptotically anti-de Sitter spacetimes}",
    eprint = "1309.1439",
    archivePrefix = "arXiv",
    primaryClass = "hep-th",
    doi = "10.1007/JHEP07(2014)086",
    journal = "JHEP",
    volume = "07",
    pages = "086",
    year = "2014"
}

@article{IHQCD1,
    author = "Gursoy, U. and Kiritsis, E.",
    title = "{Exploring improved holographic theories for QCD: Part I}",
    eprint = "0707.1324",
    archivePrefix = "arXiv",
    primaryClass = "hep-th",
    reportNumber = "CPHT-RR027-0507",
    doi = "10.1088/1126-6708/2008/02/032",
    journal = "JHEP",
    volume = "02",
    pages = "032",
    year = "2008"
}

@article{IHQCD2,
    author = "Gursoy, U. and Kiritsis, E. and Nitti, F.",
    title = "{Exploring improved holographic theories for QCD: Part II}",
    eprint = "0707.1349",
    archivePrefix = "arXiv",
    primaryClass = "hep-th",
    reportNumber = "CPHT-RR028-0507",
    doi = "10.1088/1126-6708/2008/02/019",
    journal = "JHEP",
    volume = "02",
    pages = "019",
    year = "2008"
}

@article{IHQCDthermo,
    author = "Gursoy, U. and Kiritsis, E. and Mazzanti, L. and Nitti, F.",
    title = "{Improved Holographic Yang-Mills at Finite Temperature: Comparison with Data}",
    eprint = "0903.2859",
    archivePrefix = "arXiv",
    primaryClass = "hep-th",
    doi = "10.1016/j.nuclphysb.2009.05.017",
    journal = "Nucl. Phys. B",
    volume = "820",
    pages = "148--177",
    year = "2009"
}

@article{GubserNellore,
    author = "Gubser, Steven S. and Nellore, Abhinav",
    title = "{Mimicking the QCD equation of state with a dual black hole}",
    eprint = "0804.0434",
    archivePrefix = "arXiv",
    primaryClass = "hep-th",
    reportNumber = "PUPT-2265",
    doi = "10.1103/PhysRevD.78.086007",
    journal = "Phys. Rev. D",
    volume = "78",
    pages = "086007",
    year = "2008"
}

@article{Bea2024,
    author = "Bea, Yago and Jimenez, Raul and Mateos, David and Liu, Shuheng and Protopapas, Pavlos and Taranc{\'o}n-{\'A}lvarez, Pedro and Tejerina-P{\'e}rez, Pablo",
    title = "{Gravitational duals from equations of state}",
    eprint = "2403.14763",
    archivePrefix = "arXiv",
    primaryClass = "hep-th",
    doi = "10.1007/JHEP07(2024)087",
    journal = "JHEP",
    volume = "07",
    pages = "087",
    year = "2024"
}

@article{PilchWarner,
    author = "Pilch, Krzysztof and Warner, Nicholas P.",
    title = "{N=2 supersymmetric RG flows and the IIB dilaton}",
    eprint = "hep-th/0004063",
    archivePrefix = "arXiv",
    reportNumber = "CITUSC-00-018, USC-00-02",
    doi = "10.1016/S0550-3213(00)00656-8",
    journal = "Nucl. Phys. B",
    volume = "594",
    pages = "209--228",
    year = "2001"
}

@article{BuchelPeetPolchinski,
    author = "Buchel, Alex and Peet, Amanda W. and Polchinski, Joseph",
    title = "{Gauge dual and noncommutative extension of an N=2 supergravity solution}",
    eprint = "hep-th/0008076",
    archivePrefix = "arXiv",
    reportNumber = "NSF-ITP-00-95",
    doi = "10.1103/PhysRevD.63.044009",
    journal = "Phys. Rev. D",
    volume = "63",
    pages = "044009",
    year = "2001"
}

@article{BuchelLiuThermo,
    author = "Buchel, Alex and Liu, James T.",
    title = "{Thermodynamics of the N=2* flow}",
    eprint = "hep-th/0305064",
    archivePrefix = "arXiv",
    reportNumber = "MCTP-03-25",
    doi = "10.1088/1126-6708/2003/11/031",
    journal = "JHEP",
    volume = "11",
    pages = "031",
    year = "2003"
}

@article{BuchelThermo,
    author = "Buchel, Alex and Deakin, Stan and Kerner, Patrick and Liu, James T.",
    title = "{Thermodynamics of the N=2* strongly coupled plasma}",
    eprint = "hep-th/0701142",
    archivePrefix = "arXiv",
    reportNumber = "UWO-TH-07-01, MCTP-07-04",
    doi = "10.1016/j.nuclphysb.2007.06.019",
    journal = "Nucl. Phys. B",
    volume = "784",
    pages = "72--102",
    year = "2007"
}

@article{BuchelHydro,
    author = "Buchel, Alex",
    title = "{N=2* hydrodynamics}",
    eprint = "hep-th/0406200",
    archivePrefix = "arXiv",
    doi = "10.1016/j.nuclphysb.2004.11.039",
    journal = "Nucl. Phys. B",
    volume = "708",
    pages = "451--466",
    year = "2005"
}

@article{BenincasaBuchelStarinets,
    author = "Benincasa, Paolo and Buchel, Alex and Starinets, Andrei O.",
    title = "{Sound waves in strongly coupled non-conformal gauge theory plasma}",
    eprint = "hep-th/0507026",
    archivePrefix = "arXiv",
    reportNumber = "UWO-TH-05-08",
    doi = "10.1016/j.nuclphysb.2005.11.005",
    journal = "Nucl. Phys. B",
    volume = "733",
    pages = "160--187",
    year = "2006"
}

@article{BuchelPagnuttiBulk,
    author = "Buchel, Alex and Pagnutti, Chris",
    title = "{Bulk viscosity of N=2* plasma}",
    eprint = "0812.3623",
    archivePrefix = "arXiv",
    primaryClass = "hep-th",
    reportNumber = "UWO-TH-08-18",
    doi = "10.1016/j.nuclphysb.2009.02.022",
    journal = "Nucl. Phys. B",
    volume = "816",
    pages = "62--72",
    year = "2009"
}

@article{BuchelRelaxation,
    author = "Buchel, Alex",
    title = "{Relaxation time of non-conformal plasma}",
    eprint = "0908.0108",
    archivePrefix = "arXiv",
    primaryClass = "hep-th",
    reportNumber = "UWO-TH-09-13",
    doi = "10.1016/j.physletb.2009.10.007",
    journal = "Phys. Lett. B",
    volume = "681",
    pages = "200--203",
    year = "2009"
}

@article{BuchelCritical,
    author = "Buchel, Alex and Pagnutti, Chris",
    title = "{Critical phenomena in N=2* plasma}",
    eprint = "1010.3359",
    archivePrefix = "arXiv",
    primaryClass = "hep-th",
    reportNumber = "UWO-TH-10-7",
    doi = "10.1103/PhysRevD.83.046004",
    journal = "Phys. Rev. D",
    volume = "83",
    pages = "046004",
    year = "2011"
}

@article{BuchelLiuUniversal,
    author = "Buchel, Alex and Liu, James T.",
    title = "{Universality of the shear viscosity in supergravity}",
    eprint = "hep-th/0311175",
    archivePrefix = "arXiv",
    reportNumber = "NSF-KITP-03-103, MCTP-03-53",
    doi = "10.1103/PhysRevLett.93.090602",
    journal = "Phys. Rev. Lett.",
    volume = "93",
    pages = "090602",
    year = "2004"
}

@article{BuchelUniversal,
    author = "Buchel, Alex",
    title = "{On universality of stress-energy tensor correlation functions in supergravity}",
    eprint = "hep-th/0408095",
    archivePrefix = "arXiv",
    doi = "10.1016/j.physletb.2005.01.052",
    journal = "Phys. Lett. B",
    volume = "609",
    pages = "392--401",
    year = "2005"
}

@article{BuchelCremoniniEarly,
    author = "Buchel, Alex and Cremonini, Sera and Early, Laura",
    title = "{Holographic transport beyond the supergravity approximation}",
    eprint = "2312.05377",
    archivePrefix = "arXiv",
    primaryClass = "hep-th",
    doi = "10.1007/JHEP04(2024)032",
    journal = "JHEP",
    volume = "04",
    pages = "032",
    year = "2024"
}

@article{BuchelGB,
    author = "Buchel, Alex",
    title = "{Holographic Gauss-Bonnet transport}",
    eprint = "2402.16109",
    archivePrefix = "arXiv",
    primaryClass = "hep-th",
    doi = "10.1016/j.physletb.2024.138666",
    journal = "Phys. Lett. B",
    volume = "853",
    pages = "138666",
    year = "2024"
}

@article{BuchelCharged,
    author = "Buchel, Alex and Cremonini, Sera and Moezzi, Mohammad and Tringas, George",
    title = "{Holographic charged transport with higher derivatives}",
    eprint = "2602.06144",
    archivePrefix = "arXiv",
    primaryClass = "hep-th",
    doi = "10.1007/JHEP08(2026)165",
    journal = "JHEP",
    volume = "08",
    pages = "165",
    year = "2026"
}

@article{BuchelUmbertMagnetic,
    author = "Buchel, Alex and Umbert, Bruno",
    title = "{QGP universality in a magnetic field?}",
    eprint = "2004.01583",
    archivePrefix = "arXiv",
    primaryClass = "hep-th",
    doi = "10.1007/JHEP06(2020)149",
    journal = "JHEP",
    volume = "06",
    pages = "149",
    year = "2020"
}

@article{ElingOz,
    author = "Eling, Christopher and Oz, Yaron",
    title = "{A Novel Formula for Bulk Viscosity from the Null Horizon Focusing Equation}",
    eprint = "1103.1657",
    archivePrefix = "arXiv",
    primaryClass = "hep-th",
    doi = "10.1007/JHEP06(2011)007",
    journal = "JHEP",
    volume = "06",
    pages = "007",
    year = "2011"
}

@article{BuchelEO,
    author = "Buchel, Alex",
    title = "{On Eling-Oz formula for the holographic bulk viscosity}",
    eprint = "1103.3733",
    archivePrefix = "arXiv",
    primaryClass = "hep-th",
    reportNumber = "UWO-TH-11-4",
    doi = "10.1007/JHEP05(2011)065",
    journal = "JHEP",
    volume = "05",
    pages = "065",
    year = "2011"
}

@article{FakeSupergravity,
    author = "Freedman, D. Z. and Nunez, Carlos and Schnabl, M. and Skenderis, K.",
    title = "{Fake supergravity and domain wall stability}",
    eprint = "hep-th/0312055",
    archivePrefix = "arXiv",
    reportNumber = "MIT-CTP-3450, ITFA-2003-55",
    doi = "10.1103/PhysRevD.69.104027",
    journal = "Phys. Rev. D",
    volume = "69",
    pages = "104027",
    year = "2004"
}

@misc{RaissiPINN,
 author={Maziar Raissi and Paris Perdikaris and George Em Karniadakis},
 title={{Physics Informed Deep Learning (Part I): Data-driven Solutions of Nonlinear Partial Differential Equations}},
 year={2017},
 howpublished={\href{https://arxiv.org/abs/1711.10561}{arXiv:1711.10561}}
}

@misc{QCDprogress,
 howpublished={Work in progress}
}

@article{HoyosPaikYaffe,
    author = "Hoyos, Carlos and Paik, Sangmin and Yaffe, Laurence G.",
    title = "{Screening in strongly coupled N=2* supersymmetric Yang-Mills plasma}",
    eprint = "1108.2053",
    archivePrefix = "arXiv",
    primaryClass = "hep-th",
    journal = "JHEP",
    volume = "10",
    pages = "062",
    year = "2011"
}

@article{BuchelQuench,
    author = "Buchel, Alex and Lehner, Luis and Myers, Robert C.",
    title = "{Thermal quenches in N=2* plasmas}",
    eprint = "1206.6785",
    archivePrefix = "arXiv",
    primaryClass = "hep-th",
    journal = "JHEP",
    volume = "08",
    pages = "049",
    year = "2012",
    doi = "10.1007/JHEP08(2012)049"
}

@article{BuchelHellerMyers,
    author = "Buchel, Alex and Heller, Michal P. and Myers, Robert C.",
    title = "{Equilibration rates in a strongly coupled nonconformal quark-gluon plasma}",
    eprint = "1503.07114",
    archivePrefix = "arXiv",
    primaryClass = "hep-th",
    journal = "Phys. Rev. Lett.",
    volume = "114",
    pages = "251601",
    year = "2015",
    doi = "10.1103/PhysRevLett.114.251601"
}

@article{BuchelRussoZarembo,
    author = "Buchel, Alex and Russo, Jorge G. and Zarembo, Konstantin",
    title = "{Rigorous Test of Non-conformal Holography: Wilson Loops in N=2* Theory}",
    eprint = "1301.1597",
    archivePrefix = "arXiv",
    primaryClass = "hep-th",
    journal = "JHEP",
    volume = "03",
    pages = "062",
    year = "2013",
    doi = "10.1007/JHEP03(2013)062"
}

@article{BuchelKarapetyan,
 author={Buchel, Alex and Karapetyan, Aleksandr},
 title={{de Sitter Vacua of Strongly Interacting QFT}},
 eprint={1702.01320},
 archivePrefix={arXiv},
 primaryClass={hep-th},
 journal={JHEP},
 volume={03},
 pages={114},
 year={2017},
 doi={10.1007/JHEP03(2017)114}
}

@book{PrinceDeepLearning,
    author = "Prince, Simon J. D.",
    title = "{Understanding Deep Learning}",
    publisher = "The MIT Press",
    year = "2023",
    url = "https://udlbook.github.io/udlbook/"
}

@article{DemircikAnisotropic,
 author = {Demircik, Tuna and Gallegos, Domingo and G{\"u}rsoy, Umut and J{\"a}rvinen, Matti and Lier, Ruben},
 title = {{Holographic transport in anisotropic plasmas}},
 eprint = {2402.12224},
 archivePrefix = {arXiv},
 primaryClass = {hep-th},
 year = {2024}
}
\end{document}